\documentclass[11pt,a4paper]{article}
\usepackage{jheppub}

\usepackage{amsmath}
\usepackage{amsfonts}
\usepackage{mathrsfs}
\usepackage{graphicx}
\usepackage{bigints}
\usepackage{braket}
\usepackage{mdframed}

\usepackage{tikz}
\usetikzlibrary{arrows.meta,decorations.pathmorphing, patterns,shapes}
\usepackage{eso-pic}

\usepackage[numbers,sort&compress]{natbib}

\makeatletter
\g@addto@macro\bfseries{\boldmath}
\makeatother

\newcommand{\eps}{\epsilon}

\def\beq{\begin{equation}}
\def\eeq{\end{equation}}
\def\bsp#1\esp{\begin{split}#1\end{split}}

\newcommand{\cA}{\begin{cal}A\end{cal}}

\newcommand{\cC}{\begin{cal}C\end{cal}}

\newcommand{\cI}{\begin{cal}I\end{cal}}

\newcommand\eulerian[2]{\genfrac{\langle}{\rangle}{0pt}{}{#1}{#2}}

\title{Multi-Regge kinematics and the scattering equations}

\author[a,b]{Claude Duhr}
\author[b]{Zhengwen Liu}

\affiliation[a]{Theoretical Physics Department, CERN, Geneva, Switzerland} 
\affiliation[b]{Center for Cosmology, Particle Physics and Phenomenology (CP3),\\
UCLouvain, 1348 Louvain-la-Neuve, Belgium}

\emailAdd{claude.duhr@cern.ch}
\emailAdd{zhengwen.liu@uclouvain.be}

\abstract{We study the solutions to the scattering equations in various quasi-multi-Regge regimes where the produced particles are ordered in rapidity. We observe that in all cases the solutions to the scattering equations admit the same hierarchy as the rapidity ordering, and we conjecture that this behaviour holds independently of the number of external particles. 
In multi-Regge limit, where the produced particles are strongly ordered in rapidity, we determine exactly all solutions to the scattering equations that contribute to the Cachazo-He-Yuan (CHY) formula for gluon scattering in this limit. When the CHY formula is localised on these solutions, it reproduces the expected factorisation of tree-level amplitudes in terms of impact factors and Lipatov vertices. We also investigate amplitudes in various quasi-MRK. While in these cases we cannot determine the solutions to the scattering equations exactly, we show that again our conjecture combined with the CHY formula implies the factorisation of the amplitude into universal buildings blocks for which we obtain a CHY-type representation.}

\preprint{ {CERN-TH-2018-234, CP3-18-61}}

\keywords{Multi-Regge kinematics, scattering amplitudes, scattering equations}

\begin{document}

\maketitle

\section{Introduction}
\label{sec:intro}

Making predictions for hadron collider experiments requires the computation of scattering amplitudes with external quarks and gluons. Even though Feynman diagrams allow one in principle to compute scattering amplitudes with an arbitrary number of external partons, in practise this approach is limited by the very rapid growth of the number of diagrams with the number of external particles. 

Over the last couple of years several new techniques to compute tree-level scattering amplitudes have been introduced. Among these new methods is the so-called Cachazo-He-Yuan (CHY) formalism, which expresses any $n$-point tree-level amplitude in Yang-Mills theory (and many other massless quantum field theories) as a multiple contour integral in the moduli space $\mathfrak{M}_{0,n}$ of Riemann spheres with $n$ marked points~\cite{Cachazo:2013hca,Cachazo:2013iea}. The cornerstone of the CHY formalism are the so-called {\it scattering equations} \cite{Cachazo:2013iaa, Cachazo:2013gna, Cachazo:2013hca, Fairlie:1972zz,Roberts:1972ggn, Gross:1987kza,Gross:1987ar, Witten:2004cp, Fairlie:2008dg},
which establish a map from the space of external kinematic data to the moduli space $\mathfrak{M}_{0,n}$. The CHY integral is then localised on the zeroes of the scattering equations. This allows one to reduce the computation of a tree-level scattering amplitude to a multi-dimensional residue computation. The scattering equations are equivalent to a set of polynomial equations~\cite{Dolan:2014ega}, and it is known that there are always $(n{-}3)!$ inequivalent solutions (up to $SL(2,\mathbb{C})$ transformations) for arbitrary kinematics. In general, it is very difficult to obtain exact solutions to the scattering equations valid for arbitrary multiplicities, and in practise it is only possible to solve the equations numerically for arbitrary kinematics (cf.~e.g., ref.\,\cite{Farrow:2018cqi,Liu:2018brz}).

The representation of a tree-level scattering amplitude as a residue integral also makes manifest the properties of the amplitude in certain singular limits. For example, it was shown that the limit in which a gluon is soft has a very natural interpretation in the CHY language in terms of a simple residue computation~\cite{Cachazo:2013hca}. As a consequence, the well-known factorisation of a tree-level gluon amplitude in the soft limit into an eikonal factor and a lower-point amplitude has a very simple derivation in the CHY framework. The analysis of the soft limit in the CHY framework was subsequently extended beyond leading order in various theories \cite{Schwab:2014xua,Afkhami-Jeddi:2014fia,Kalousios:2014uva,Zlotnikov:2014sva,Cachazo:2015ksa,Zlotnikov:2017ahq,Chakrabarti:2017zmh}. Similarly, the factorisation of a gluon amplitude in the collinear limit was derived from the CHY representation of the amplitude in ref.\,\cite{Nandan:2016ohb}. 

The aim of this paper is to show that the CHY formalism also presents a very natural framework to study other kinematical limits of tree-level amplitudes in gauge theory, namely the so-called Regge limits of a 2-to-$(n{-}2)$ scattering where the produced particles are ordered in rapidity. Of particular interest in this context is multi-Regge kinematics (MRK) where all produced gluons are strongly ordered in rapidity while having comparable transverse momenta. It is expected that in MRK any tree-level amplitude in Yang-Mills theory factorises into a product of universal building blocks known as impact factors and Lipatov vertices~\cite{Kuraev:1976ge,Lipatov:1976zz,Lipatov:1991nf}, connected by $t$-channel propagators. By relaxing the strong ordering among the rapidities of the produced particles, one can define a tower of new kinematical limits, known as quasi-multi-Regge kinematics (QMRK). In each such limit the amplitude is again expected to factorise into a product of universal building blocks which are multi-particle generalisations of the impact factors and Lipatov vertices~\cite{Fadin:1989kf}. While consistent with all known results for tree-level amplitudes, so far the factorisation of colour-ordered gluon helicity amplitudes in (quasi-)MRK remains partly conjectural and has only been proven to hold for arbitrary multiplicities for the simplest non-zero helicity configurations~\cite{DelDuca:1995zy}. 

One of the goals of this paper is to use the CHY representation of gluon amplitudes to shed light on the factorisation in QMRK. Through a numerical study of the solutions to the scattering equations in various quasi-multi-Regge limits, we observe that in all cases the solutions in the limit present the same hierarchy as the rapidity ordering that defines the limit (if the $SL(2,\mathbb{C})$ redundancy is fixed in a certain way). We conjecture that this feature holds in general, independently of the helicity configuration and the number of external legs. While we do currently not have a proof of our conjecture, we show that when combined with the CHY representation of gluon amplitudes it reproduces the expected factorisation in certain quasi-multi-Regge limits. This gives very strong support to the validity of our conjecture. At the same time, our results reveal the CHY origin of the Regge factorisation of tree-level amplitudes, and that it is possible to derive the factorisation of the amplitude directly from the CHY representation. As a byproduct of our analysis, we obtain CHY-type representations for the generalised impact factors and Lipatov vertices that appear in QMRK for an arbitrary number of produced particles. Finally, we show that in MRK our conjecture implies that the (four-dimensional) scattering equations have a unique solution in each `helicity configuration', and we determine this solution explicitly for arbitrary multiplicities.

The paper is organised as follows: In Section~\ref{sec-Rev-CHY-MRK} we give a short review of the background on the CHY formalism and Regge kinematics needed in the remainder of the paper. In Section~\ref{sec-SEs-MRK} we present our main conjecture about the behaviour of the solutions to the scattering equations in QMRK, and in Section~\ref{sec-MRK} we apply it to explicitly solve the scattering equations in MRK and to derive the factorisation of the amplitude in this limit. Finally, in Section~\ref{sec-QMRK} we extend the analysis to various quasi-multi-Regge limits and we obtain CHY-type representations for the generalised impact factors and Lipatov vertices. We include several appendices where we discuss certain aspects omitted throughout the main text.


\section{Review of the Cachazo-He-Yuan formalism and Regge kinematics}
\label{sec-Rev-CHY-MRK}

In order to fix our notations and conventions, we give in this section a brief review on the Cachazo-He-Yuan (CHY) formalism and the scattering equations, as well as the behaviour of tree-level scattering amplitudes in various Regge limits.

\subsection{The Cachazo-He-Yuan formalism}

Our main objects of interest are tree-level scattering amplitudes for massless particles in an $SU(N)$ gauge theory. We focus here on a scattering of $n$ massless gauge bosons with momenta $k_a$, $1\le a\le n$, that are on shell, $k_a^2=0$, and satisfy  momentum conservation, $\sum_{a=1}^nk_a^\mu = 0$. We always assume that we are working with colour-ordered amplitudes associated with a given cyclic ordering of the external particles. The functional form of the colour-ordered amplitudes is determined by the helicities of the gluons. If at most one gluon has a negative  helicity, then the amplitude vanishes. The first non-zero amplitudes are then those where exactly two gluons have a negative (positive) helicity, and they are refereed to as maximal helicity-violating (MHV) amplitudes (or $\overline{\text{MHV}}$ amplitudes in the case of two positive helicities). In general, an amplitude involving $k\ge 2$ negative helicities is referred to as a (next-to-)$^{k-2}$MHV (N$^{k-2}$MHV) amplitude.

There are various different ways to compute tree-level amplitudes. In this paper we are mostly interested in the CHY formalism, where any $n$-point tree-level amplitude is expressed as a multiple integral over the moduli space $\mathfrak{M}_{0,n}$ of Riemann spheres with $n$ marked points~\cite{Cachazo:2013hca, Cachazo:2013iea},
\begin{align}\label{CHY-1}
  {\cal A}_n  \,=\, \int {\prod_{a=1}^{n} d\sigma_a \over \operatorname{vol}\,{SL}(2,{\mathbb C})}\,
  {\prod}'_a \delta\left( f_a \right)\, {\cal I}_n\,.
\end{align}
Since tree-level amplitudes are rational functions, the integrand is completely localised by the $\delta$-functions, and we define
\beq\label{eq:delta_functions}
{\prod}'_a \delta\left( f_a \right) \,\equiv\, (\sigma_{rp}\sigma_{pq}\sigma_{qr} )\prod_{a\ne r,p,q} \delta\left( f_a \right)\,,
\eeq
where $\sigma_{ab} \equiv \sigma_{a} {-} \sigma_{b}$. The arguments of the $\delta$-functions are the {\it scattering equations}, which establish a map from a configuration of $n$ massless momenta to the moduli space $\mathfrak{M}_{0,n}$
\begin{align}\label{SE-chy-1}
  f_a \,=\, \sum_{b\neq a} {k_a\cdot k_b \over \sigma_{a}  {-} \sigma_{b}} \,=\, 0\,,
  \qquad a=1, 2, \ldots, n\,.
\end{align}
The last ingredient of eq.\,\eqref{CHY-1} is the integrand ${\cal I}_n$ which contains the information on the dynamics of the theory. The precise form of the integrand $\cI_n$ in the case of gluon scattering is irrelevant for the purpose of this paper, so we do not show it here explicitly.

Let us make some comments about the properties of the CHY representation of scattering amplitudes in eq.\,\eqref{CHY-1}.
First, the integral in eq.\,\eqref{CHY-1} is invariant under M\"{o}bius transformations for $SL(2,\mathbb{C})$ and the integration measure transforms covariantly.
We can thus fix three of the variables {(e.g.~$\sigma_1=0$, $\sigma_2\to\infty$ and $\sigma_3=1$)} and use the $(n{-}3)$ $\delta$-functions in eq.\,\eqref{eq:delta_functions} to completely localise the remaining integration variables. Second, it has been proven that the number of independent solutions of the scattering equations in eq.\,\eqref{SE-chy-1} is $(n{-}3)!$ \cite{Cachazo:2013gna}. The scattering amplitude $\cA_n$ is then obtained by summing over all independent solutions.

Equation~\eqref{CHY-1} is valid for scattering amplitudes in any space-time dimension. We now review an alternative formulation of eq.\,\eqref{CHY-1} which is restricted to four space-time dimensions. In four dimensions we can write any massless momentum as a product of two spinors with opposite chirality, $k_a^{\alpha\dot\alpha} = \lambda_a^{\alpha}\tilde{\lambda}_a^{\dot{\alpha}}$. The two spinors are complex conjugate to each other if the momentum is real. Since the scattering equations in eq.\,\eqref{SE-chy-1} establish a map from a configuration of momenta to $\mathfrak{M}_{0,n}$, it is natural to ask if the scattering equations themselves can be split into different `helicity sectors'\footnote{We emphasise that the helicity sector of the scattering equations is different from the helicity configuration of the external particles.} involving only spinors with definite chirality.
It was shown that this is indeed possible (cf.~ref.\,\cite{He:2016vfi} as well as Appendix \ref{app:4dSE}). Each helicity sector is characterised by an integer $k$, with $2\le k \le n{-}2$, and in the $k$ sector the scattering equations can be reduced to the following spinor-valued equations \cite{Geyer:2014fka},
\beq\bsp\label{SE-4d-twistor}
  \bar{\cal E}^{\dot\alpha}_I \,&=\,  \tilde\lambda_I^{\dot\alpha}  - \sum_{i\in\mathfrak{P}} {t_I t_i \over \sigma_{I} {-} \sigma_{i}}\tilde\lambda_i^{\dot\alpha} \,=\, 0,
  ~~I\in\mathfrak{N}\,,\\
  {\cal E}^{\alpha}_i \,&=\,  \lambda_i^{\alpha} - \sum_{I\in\mathfrak{N}} {t_i t_I \over \sigma_{i} {-} \sigma_{I}}\lambda^\alpha_I \,=\, 0,
  ~~i\in\mathfrak{P},
\esp\eeq
where $\mathfrak{N}$ is a subset of $\{1,\ldots,n\}$ with length $k$ and $\mathfrak{P}$ is its complement.
We refer to eq.\,\eqref{SE-4d-twistor} as the \emph{four-dimensional scattering equations}. They were originally derived from the four-dimensional ambitwistor string model in ref.\,\cite{Geyer:2014fka}.
The tree-level amplitude for $n$-gluon scattering then takes the form~\cite{Geyer:2014fka},
\begin{align}\label{amp-YM-4d}
   {\mathscr A}_n(1, 2, \ldots, n) & \,=\, \delta^4\left(\sum_{a=1}^{n} k_a^\mu \right)\,\cA_n(1, 2, \ldots, n)\\
\nonumber   &\,=\,\bigintsss
  {\prod_{a=1}^{n} d^2\sigma_a \over \text{vol}\,{GL}(2,\mathbb{C})}\,
  \prod_{I\in\mathfrak{N}} 
  \delta^{2}\big( \bar{\cal E}^{\dot\alpha}_I \big)
  \prod_{i\in\mathfrak{P}} \delta^2\big( {\cal E}^{\alpha}_i \big)\,\cI_n(1, 2, \ldots, n),
\end{align}
where $(a\,b) \equiv (\sigma_a {-} \sigma_b)/(t_a t_b)$ and $d^2\sigma_a = d\sigma_a dt_a/t_a^3$.
{
By abuse of language, we refer here to both the $D$-dimensional and four-dimensional formulas in eqs.~\eqref{CHY-1} and~\eqref{amp-YM-4d} for tree-level amplitudes as \emph{Cachazo-He-Yuan (CHY) formulas}, even though eq.~\eqref{amp-YM-4d} was originally derived from  ambitwistor string theory by Geyer, Lipstein and Mason.}
An important property is that only the equations of sector $|\mathfrak{N}| = k$ contribute to N$^{k-2}$MHV amplitudes.
When we use $\mathfrak{N}$ and $\mathfrak{P}$ to collect the labels of gluons with negative and positive helicity respectively, simplifications can occur and ${\cal I}_n$ takes the following simple form
\beq\label{eq:YM_integrand_4d}
\cI_n(1,2,\ldots, n) \,=\,  {1\over (1\,2)(2\,3) \cdots (n\,1)}\,.
 \eeq
We also note that we can use the $GL(2,\mathbb{C})$ invariance of eq.\,\eqref{amp-YM-4d} to fix four variables, and we can eliminate four of the $\delta$-functions and identify them with the delta-function expressing momentum conservation~\cite{Geyer:2014fka}, e.g., in the case $\{1,2\}\subseteq \mathfrak{N}$,
\begin{align}\label{SEs-12-momentum-conserv}
  \delta^2\big( \bar{\cal E}^{\dot\alpha}_1 \big)  \delta^2\big( \bar{\cal E}^{\dot\alpha}_2 \big)
  \,=\, \braket{1\,2}^2\, \delta^4\left(\sum_{a=1}^{n} k_a^\mu \right)\,.
\end{align}
In this paper we will mostly work with the four-dimensional versions in eqs.~\eqref{SE-4d-twistor} and~\eqref{amp-YM-4d} to compute gluons amplitudes, because it has a simpler structure than the $D$-dimensional version in eq.~\eqref{CHY-1} and is more suited to study helicity amplitudes.
Without loss of generality, here we always use the convention where $\{1,2\}\subseteq \mathfrak{N}$.


\subsection{Multi-Regge kinematics}

Multi-Regge kinematics (MRK) is defined as a $2$-to-$(n{-}2)$ scattering of partons (here we only consider gluons) where the produced particles in the final state are strongly ordered in rapidity while having comparable transverse momenta,
\begin{align}\label{MRK-rapidity}
  y_3 \gg y_4 \gg \cdots \gg y_n \textrm{~~and~~} |k_3^\perp| \simeq|k_4^\perp|\simeq\ldots\simeq |k_n^\perp|\,,
\end{align}
and we introduce the complexified transverse momenta $k_a^{\perp}=k_a^1 + i k_a^2$. 
In terms of lightcone coordinates $k_a=(k_a^+,k_a^-; k_a^{\perp})$, with $k_a^{\pm}=k_a^0\pm k_a^3$, the strong ordering in rapidities is equivalent to a strong ordering in lightcone $+$-components,
\begin{align}\label{MRK-motenta-order}
  k_3^+ \gg k_4^+ \gg \cdots \gg k_n^+\,.
\end{align}
In the following we always assume that we are working in the center-of-mass frame where the two incoming gluons are back-to-back on the $z$-axis,
\begin{align}
  k_1 \,=\, (0, -{\kappa}; 0), \quad k_2 \,=\, (-{\kappa}, 0; 0)\,,\quad \kappa\equiv \sqrt{s}\,,
\end{align}
where $s$ is the squared center-of-mass energy.

\begin{figure}[t]
  \centering
  \begin{tikzpicture}[scale=1]
  
  \newcommand{\nicearrow}{-{Latex[length=7mm, width=1.5mm]}}
  
  \fill[blue,opacity=0.25] (0,0) circle (0.25);
  \draw[line width=1.0pt] (0,0) circle (0.25);
  
  \draw[\nicearrow, line width=0.9pt] (170:2.5)--(170:1.0);
  \draw[-, line width=0.9pt] (170:0.25) -- (170:2.5) node[left] {\large$k_2$};

  \draw[\nicearrow, line width=0.9pt] (10:2.5)--(10:1.0);
  \draw[-, line width=0.9pt] (10:0.23) -- (10:2.5) node[right] {\large$k_3$};

  \fill[yshift=-1.7cm, green, opacity=0.20] (0,0) circle (0.25);
  \draw[yshift=-1.7cm, line width=1.0pt] (0,0) circle (0.25);
  
  \draw[yshift=-1.7cm, \nicearrow, line width=0.9pt] (0:2.5)--(0:1.0);
  \draw[yshift=-1.7cm, -, line width=0.9pt] (0:0.25) -- (0:2.5) node[right] {\large$k_4$};

  \draw[yshift=-0.25cm, decorate,decoration=zigzag, line width=0.9pt] (0,0) -- (0,-1.2);
  \draw[yshift=-0.5cm, xshift=0.3cm, \nicearrow, line width=0.8pt] (0,0) -- (0,-1.0);
  \fill[yshift=-0.8cm, xshift=0.3cm] (0,0)circle (0pt) node[right] {\large$q_4$};
  
  
   \draw[yshift=-1.95cm, decorate,decoration=zigzag, line width=0.9pt] (0,0) -- (0,-0.75);
   
   \draw[yshift=-2.77cm, dashed, line width=0.9pt] (0,0) -- (0,-0.75);
   
   \draw[yshift=-4.25cm, decorate,decoration=zigzag, line width=0.9pt] (0,0) -- (0,0.75);
   
  \fill[yshift=-4.5cm, green, opacity=0.20] (0,0) circle (0.25);
  \draw[yshift=-4.5cm, line width=1.0pt] (0,0) circle (0.25);
  
  \draw[yshift=-4.5cm, \nicearrow, line width=0.9pt] (0:2.5)--(0:1.0);
  \draw[yshift=-4.5cm, -, line width=0.9pt] (0:0.25) -- (0:2.5) node[right] {\large$k_{n-1}$};
  
 \draw[yshift=-4.75cm, decorate,decoration=zigzag, line width=0.9pt] (0,0) -- (0,-1.2);
  \draw[yshift=-5.0cm, xshift=0.3cm, \nicearrow, line width=0.8pt] (0,0) -- (0,-1.0);
  \fill[yshift=-5.3cm, xshift=0.3cm] (0,0)circle (0pt) node[right] {\large$q_n$};
  
  \fill[yshift=-6.2cm, blue,opacity=0.25] (0,0) circle (0.25);
  \draw[yshift=-6.2cm, line width=1.0pt] (0,0) circle (0.25);
  
  \draw[yshift=-6.2cm, \nicearrow, line width=0.9pt] (190:2.5)--(190:1.0);
  \draw[yshift=-6.2cm, -, line width=0.9pt] (190:0.25) -- (190:2.5) node[left] {\large$k_1$};

  \draw[yshift=-6.2cm, \nicearrow, line width=0.9pt] (-10:2.5)--(-10:1.0);
  \draw[yshift=-6.2cm, -, line width=0.9pt] (-10:0.23) -- (-10:2.5) node[right] {\large$k_n$};
  
\end{tikzpicture}
\caption{\label{fig:mrk}The factorised form of a tree-level amplitude in multi-Regge kinematics.}
\end{figure}
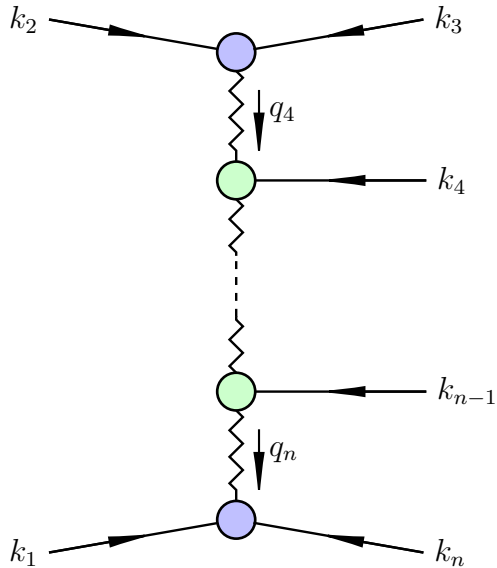

It is conjectured that in MRK every gluon amplitude factorises into a set of universal building blocks describing the emission of gluons along a $t$-channel ladder (see fig.~\ref{fig:mrk}). More precisely, one has
\begin{align}\label{amp-MRK-main}
  \!\!\!{\cal A}_n(1,\!\ldots\!,n)&\simeq s\,C(2;3) {-1 \over |{q_{4}^\perp}|^2} V(q_{4}; 4; q_{5}) \cdots
  {-1 \over |{q_{n-1}^\perp}|^2} V(q_{n-1}; n{-}1; q_{n}) {-1 \over |{q_{n}^\perp}|^2} C(1; n)\,,
\end{align}
where $q_{a} = \sum_{b=2}^{a-1} k_b$ with $4\le a \le n$ are the momenta exchanged in the $t$-channel. We use the `$\simeq$' sign to denote equality up to terms that are power-suppressed in the limit. The quantities that appear in the right-hand side of eq.\,\eqref{amp-MRK-main} are the \emph{impact factors}~\cite{Kuraev:1976ge, DelDuca:1995zy},
\beq\bsp\label{eq:impact_factors}
  C(2^+;3^+) \,&=\, C(2^-;3^-) \,=\, 0 \,,\\
  C(2^-;3^+)  \,&=\, C(2^+;3^-) \,=\,1 \,,\\
  C(1^+;n^+) \,&=\, C(1^-;n^-) \,=\, 0 \,,\\
  C(1^-;n^+)  \,&=\, C(1^+;n^-)^\ast \,=\, {({k_n^\perp})^\ast \over {k_n^\perp}}\,,
\esp\eeq
and the \emph{Lipatov vertices}~\cite{DelDuca:1995zy, Lipatov:1976zz, Lipatov:1991nf},
\beq\label{eq:Lipatov_vertex}
 V\big( q_{a}; a^+; q_{a+1} \big) \,=\, V\big( q_{a}; a^-; q_{a+1} \big)^\ast  \,=\,  {(q_{a}^\perp)^\ast\, q_{a+1}^\perp  \over k_a^\perp}\,.
 \eeq
Helicity is conserved by the impact factors, which forces some of the helicity combinations in eq.\,\eqref{eq:impact_factors} to vanish. The impact factors and Lipatov vertices appearing in MRK are entirely determined by the four- and five-point amplitudes. Since there are no non-MHV helicity configurations for four and five particles, we see that in MRK any amplitude is determined by MHV-type building blocks, independently of the helicity configuration. Although consistent with all explicit results for tree-level amplitudes,  eq.\,\eqref{amp-MRK-main} was only rigorously proven for arbitrary multiplicity for the simplest helicity configurations~\cite{DelDuca:1995zy}. One of the aims of this paper is to explore in how far the CHY formalism can be used to shed light on the factorisation of scattering amplitudes in Regge kinematics.

Starting from the multi-Regge limit in eq.\,\eqref{MRK-rapidity}, one can define a tower of new kinematic regimes, called quasi-multi-Regge kinematics (QMRK), by relaxing the hierarchy among some of the rapidities of the produced particles, e.g.,
 \begin{align}\label{QMRK-rapidity}
  y_3 \simeq \cdots \simeq y_k\gg y_{k+1}\simeq\cdots\simeq y_{n-1}\gg y_n \,,
\end{align}
while all transverse components are of the same size.
In QMRK the amplitude is conjectured to factorise in a way very similar to MRK in eq.\,\eqref{amp-MRK-main}. For example, in the limit in eq.\,\eqref{QMRK-rapidity}, the amplitude is expected to factorise as follows:
\begin{align}\label{amp-QMRK-main}
  {\cal A}_n(1,\ldots,n) \,&\simeq\, s\,C(2;3,\cdots,k)\, {-1 \over |{q_{k+1}^\perp}|^2}\, V\big(q_{k+1}; k{+}1,\cdots n{-}1; q_{n}\big) 
  {-1 \over |{q_{n}^\perp}|^2} C(1; n)\,.
\end{align}
Impact factors for the production of up to four particles have been computed in ref.~\cite{DelDuca:1995zy,DelDuca:1999iql} from helicity amplitudes, and  in ref.~\cite{Antonov:2004hh} using the effective action for high-energy processes in QCD~\cite{Lipatov:1995pn}. The Lipatov vertices are known for the emission of up to four particles emitted in the center~\cite{DelDuca:1995zy,DelDuca:1999iql,Duhr:2009uxa}. 

Let us conclude this section with a comment on how we implement the (quasi-)multi-Regge limit on the amplitudes. We start by introducing a small parameter $\epsilon$, and we rescale the final state momenta according to~\cite{Prygarin:2011gd}
\beq\label{QMRK-parameterization}
k_a \,=\, (k_a^+,k_a^-; k_a^\perp) \,\rightarrow\, \big(k_a^+\epsilon^{\alpha_a},k_a^-\epsilon^{-\alpha_a}; k_a^\perp\big)\,,\quad 3\le a\le n\,,
\eeq
where the $\alpha_a$ are rational numbers. We can then adjust the values of the $\alpha_a$ such as to reproduce the hierachy in a given quasi-multi-Regge limit. For example, the choice $\alpha_a = a-\frac{n-3}{2}$, reproduces the hierarchy of MRK in eq.\,\eqref{MRK-rapidity}. Similarly, the choice $\alpha_a=-1$ ($3\le a \le k$), $\alpha_b=0$ with $k< b < n$ and $\alpha_n=1$ allows one to approach the quasi-multi-Regge limit of eq.\,\eqref{QMRK-rapidity}.
In four dimensions, using spinor-helicity variables, the rescaling of the lightcone coordinates immediately translates into a rescaling of the components of the spinors,
\begin{align}\label{MRK-spinor}
\begin{aligned}
  \lambda_1 \,=\, - \tilde\lambda_1 \,=\, 
  \left(\begin{matrix}  0  \\  \sqrt{\kappa} \end{matrix}\right),
  &\quad
  \lambda_2 \,=\, -\tilde\lambda_2 \,=\, 
  \left(\begin{matrix}  \sqrt{\kappa}  \\  0  \end{matrix}\right),
  \\
  \lambda_a \,=\, 
  \left(\begin{array}{l} \sqrt{k_a^+}\,\epsilon^{\alpha_a/2}  \\  \sqrt{k_a^-}e^{i\phi_a}\,\epsilon^{-\alpha_a/2}  \end{array}\right),
  &\quad
  \tilde\lambda_a \,=\, 
  \left(\begin{array}{l} \sqrt{k_a^+}\,\epsilon^{\alpha_a/2}  \\  \sqrt{k_a^-}e^{-i\phi_a}\,\epsilon^{-\alpha_a/2}  \end{array}\right),
  \quad 3\le a\le n\,,
\end{aligned}
\end{align}
where the phase is given by $e^{i\phi_a}=\sqrt{k_a^\perp/(k_a^\perp)^\ast}$. In the next section we will use this parametrisation of the quasi-multi-Regge limit to study the behaviour of the scattering equations in the limit.


\section{The scattering equations in Regge kinematics}
\label{sec-SEs-MRK}

\subsection{A warm up: the MHV sector in Regge kinematics}
\label{sec:warm_up}
The aim of this section is to study the behaviour of the solutions of the scattering equations in various quasi-multi-Regge limits. Before we present in the next subsection a concise conjecture on the behaviour of the solutions in these limits, we find it instructive to analyse in detail the MHV sector, $k=2$. We do this for two reasons. First, we have argued in the previous section that in MRK any amplitude is determined by MHV-type building blocks, so the MHV sector should capture a lot of information on the multi-Regge limit. Second, while in general it is very hard, if not impossible, to find exact analytic solutions to the scattering equations, in the MHV sector one can solve the equations exactly and study their behaviour analytically.
In particular, there is a unique solution in the MHV sector (up to ${SL}(2,{\mathbb C})$ transformations) \cite{Fairlie:1972zz, Roberts:1972ggn, Fairlie:2008dg} (see also refs.~\cite{Weinzierl:2014vwa, Dolan:2014ega}). If we use the ${SL}(2,{\mathbb C})$ redundancy to fix three of the $\sigma_a$, e.g.,
\beq\label{gauge-fixing-d-se}
\sigma_1 \,=\, 0 \,,\qquad \sigma_2 \,=\, \infty\,,\qquad \sigma_3 \,=\, {k_3^+ \over k_3^\perp}\,,
\eeq
then this solution can be written as,
\begin{align}\label{MHV-sol}
  \sigma_a \,=\, {{k_a^+} \over {k_a^\perp}}~~~\text{for}~~~4 \le a \le n\,.
\end{align}
We refer to this particular solution as the \emph{MHV solution}.
It is easy to check that the complex conjugate of this solution is the solution in the $\overline{\textrm{MHV}}$ sector $k=n{-}2$.

Let us now analyse what the MHV solution becomes in a quasi-multi-Regge limit.
From eqs.\,\eqref{gauge-fixing-d-se} and \eqref{MHV-sol} it is easy to see that for the MHV solution (and also for its complex conjugate) the $\sigma_a$, $a>2$, are of the same order of magnitude as the corresponding lightcone coordinates $k_a^+$, i.e.~$\sigma_a = {\cal O}(k_a^+)$. More precisely, we see that for the MHV solution in eqs.\,\eqref{gauge-fixing-d-se} and \eqref{MHV-sol} we have
\beq\label{eq:conj_MHV}
\operatorname{Re}(\sigma_a) = {\cal O}(k_a^+) \textrm{~~and~~}\operatorname{Im}(\sigma_a) = {\cal O}(k_a^+), \quad 3\le a\le n\,.
\eeq
In other words, in QMRK the MHV solution admits the same strong ordering as the rapidities.
In particular, in MRK the MHV solution admits a hierarchy very reminiscent of the MRK hierarchy in eq.\,\eqref{MRK-rapidity},
\begin{align}\label{MHV-sol-ordering}
  |\operatorname{Re}(\sigma_3)| \gg \cdots \gg |\operatorname{Re}(\sigma_n)|
  \quad\text{and}\quad
  |\operatorname{Im}(\sigma_3)| \gg \cdots \gg |\operatorname{Im}(\sigma_n)|\,.
\end{align}

Let us also look at the MHV solution of the four-dimensional scattering equations in eq.\,\eqref{SE-4d-twistor}. They do not only depend on the variables $\sigma_a$, but also on $t_a$. For concreteness, since we use the convention that $\{1,2\}\subseteq\mathfrak{N}$, in the MHV case we have $\mathfrak{N}=\{1, 2\}$.
We can then use $GL(2,{\mathbb C})$ redundancy to fix four variables as follows:
\begin{align}\label{gauge-fixing-4d-se}
  \sigma_1 \,=\,  0,  \quad t_2 \,=\, \sigma_2 \,=\, \infty, \quad t_1 \,=\, -1\,.
\end{align}
In addition, we can eliminate four equations using eq.\,\eqref{SEs-12-momentum-conserv}. 
We can then solve the remaining $2n{-}4$ equations, and the unique solution 
takes the form
\begin{align}\label{eq:MHV_solution}
  \sigma_a \,=\, {{k_a^+} \over {k_a^\perp}}, \qquad t_a \,=\, - {\sqrt{k_a^+ \over \kappa}}\,, \quad {a \ge 3}\,.
\end{align}
We see that, as expected, the $\sigma_a$ behave again according to eq.\,\eqref{eq:conj_MHV}. The variables $t_i$ instead behave like
\beq\label{eq:conj_t_MHV}
t_a = {\cal O}\left(\sqrt{k_a^+/\kappa}\right)\,, \quad {a \ge 3} \,.
\eeq

\subsection{The main conjectures}

Our analysis in the previous section shows that in every quasi-multi-Regge limit the MHV solution admits the same hierarchy as the rapidities of the produced particles. The purpose of this section is to extend this observation to the solutions of the scattering equations in other sectors. More precisely, we propose the following conjecture:

\begin{mdframed}
{\bf Main Conjecture: }In QMRK, if the $SL(2,\mathbb{C})$ redundancy is fixed according to eq.\,\eqref{gauge-fixing-d-se}, then all solutions of the $D$-dimensional scattering equations satisfy the same hierarchy as the ordering of the rapidities that defines the QMRK. This holds for all $\sigma_a$ solutions of four-dimensional scattering equations after fixing the $GL(2,\mathbb{C})$ redundancy according to eq.\,\eqref{gauge-fixing-4d-se}.
A particular example is that in MRK, we conjecture
\begin{align}\label{}
\begin{aligned}
  &|\operatorname{Re}(\sigma_3)| \gg |\operatorname{Re}(\sigma_4)| \gg \cdots \gg |\operatorname{Re}(\sigma_n)|
  \\
  \text{and}\quad
  &|\operatorname{Im}(\sigma_3)| \gg |\operatorname{Im}(\sigma_4)| \gg \cdots \gg |\operatorname{Im}(\sigma_n)|.
\end{aligned}
\end{align}
Furthermore, in QMRK for the four-dimensional scattering equations, the two sets of solutions $(t_i)_{i\in \mathfrak{P}}$ and $(t_I)_{I\in \mathfrak{N}}$ are individually ordered according to the same hierarchy as the rapidities (though there may not be any ordering between elements belonging to different sets).
\end{mdframed}

We have performed a detailed numerical analysis of the scattering equations for a set of external momenta which approach a quasi-multi-Regge regime, both in their $D$ and their four-dimensional guises.  
We can approach a given quasi-multi-Regge limit numerically by choosing an appropriate numerical hierarchy between the lightcone components of the external momenta using eq.\,\eqref{QMRK-parameterization}. 
We have solved the $D$-dimensional scattering equations up to eight points, and the four-dimensional scattering equations for all helicity sectors up to seven external particles. All results of the numerical analyses agree with our conjecture.

We can formulate our conjecture more sharply and give the precise behaviour of the solutions in QMRL. In order to do so, it is convenient to discuss separately the four- and $D$-dimensional cases.
Let us start by discussing the results of the numerical analysis for the four-dimensional scattering equations in eq.\,\eqref{SE-4d-twistor}.
We observe that in all cases, including for the non-MHV sectors $k>2$, all solutions of the scattering equations in eq.\,\eqref{4dSE-lightcone} behave in a similar way as the MHV solution in eqs.~\eqref{eq:conj_MHV} and~\eqref{eq:conj_t_MHV}. Based on this study, we are led to state the following conjecture:
\begin{mdframed}
{\bf Conjecture 1 (four-dimensional version): }In QMRK, if $\{3,n\}\subseteq\mathfrak{P}$ and the $GL(2,\mathbb{C})$ redundancy is fixed according to eq.\,\eqref{gauge-fixing-4d-se}, then all solutions of the four-dimensional scattering equations behave as, with $3\le a\le n$,
\begin{align}\label{eq:main_conj}
  \operatorname{Re}(\sigma_a) \,=\, {\cal O}\left( {k_a^+} \right),\quad 
  \operatorname{Im}(\sigma_a) \,=\, {\cal O}\left( {k_a^+} \right), \quad 
  t_a \,=\, {\cal O}\left(\!\!\sqrt{k_a^+\,\kappa^{-h_a}}\right)\,,
\end{align}
where $h_a=1$ when $a\in\mathfrak{P}$, otherwise $h_a=-1$.
\end{mdframed}
We stress that the previous conjecture only holds in the case $\{3,n\}\subseteq\mathfrak{P}$, and the solutions $t_a$ have a different behaviour in QMRK in other cases. We have also performed a numerical analysis of the behaviour in these other cases, but since only the case $\{3,n\}\subseteq\mathfrak{P}$ will be relevant for gluon scattering in MRK when we use eq.~\eqref{amp-YM-4d}, we only present this case in the main text and we defer a discussion of the remaining cases to Appendix~\ref{app:conjecture}. Note that the helicity sectors $\mathfrak{P}$ and $\mathfrak{N}$ are not directly related to the helicities carried by the external particles, and we can use the scattering equations in eq.~\eqref{SE-4d-twistor} with $\{3,n\}\subseteq\mathfrak{P}$ even if particles 3 and/or $n$ have negative helicity.

Since for given a set of kinematic data the solutions $\sigma_a$ of the four-dimensional scattering equations also satisfy the $D$-dimensional scattering equations in eq.\,\eqref{SE-chy-1} (see Appendix~\ref{app:4dSE}), it is natural to conjecture that the solutions to the $D$-dimensional scattering equations in eq.\,\eqref{SE-chy-1} have the same behaviour in QMRK as their four-dimensional counterparts. More precisely, we have the following conjecture:
\begin{mdframed}
{\bf Conjecture 2 ($D$-dimensional version):} In QMRK, if the $SL(2,\mathbb{C})$ redundancy is fixed according to eq.\,\eqref{gauge-fixing-d-se}, then all solutions of the $D$-dimensional scattering equations satisfy
\begin{align}
  \operatorname{Re}(\sigma_a) \,=\, {\cal O}\left( {k_a^+} \right),\quad 
  \operatorname{Im}(\sigma_a) \,=\, {\cal O}\left( {k_a^+} \right),  \quad 4\le a\le n.
\end{align}
\end{mdframed}

These conjectures are among the main results of our paper. 
Although we do currently not have any proof of our conjectures, we find it remarkable that in all cases we have investigated the solutions to the scattering equations display this universal asymptotic behaviour, independently of the helicity sector. In the remainder of this paper we explore the consequences of our conjectures. In particular, we will show that they imply the factorisation of tree-level amplitudes in quasi-multi-Regge limits when combined with the CHY representation of tree-level amplitudes. This result is important for two reasons: First, it gives support to our conjectures, because it shows that they allow us to recover expected behaviour of tree-level amplitudes in quasi-multi-Regge limits. Second, it reveals the CHY origin of Regge factorisation. Since the scattering equations are generic and independent of the theory, it would be interesting to investigate if our conjecture implies universal Regge factorisation properties not only for gauge theories, but also for other theories. We leave this investigation for future work. Finally, let us make a comment at this point. Equation~\eqref{eq:main_conj} implies that the variables $\sigma_a$ scale in every helicity sector in the same way as in the MHV sector, cf.~eq.~\eqref{eq:MHV_solution}. The scaling of the variables $t_a$ in eq.~\eqref{eq:main_conj}, however, depends on the `helicity configuration'. 
This is natural in the sense that, even away from Regge kinematics the solutions $\sigma_a$ are independent of the ways of dividing $n$ labels into two subsets $\mathfrak{N}$ and $\mathfrak{P}$, $|\mathfrak{N}|=k$, $2\le k \le n{-}2$, and these solutions form a subset of the $(n{-}3)!$ solutions to the $D$-dimensional scattering equations.
Very differently, the variables $t_a$ are specific to the four-dimensional scattering equations and take different values in different configurations for given $k$.


\section{Multi-Regge factorisation from the CHY formalism}
\label{sec-MRK}
The goal of this section is to apply Conjecture 1 to the case of multi-Regge kinematics where the produced particles are strongly ordered in rapidity, cf. eq.\,\eqref{MRK-rapidity}. In Section~\ref{sec-Rev-CHY-MRK} we have reviewed that in MRK the amplitude is conjectured to take a simple factorised form, cf.~eq.\,\eqref{amp-MRK-main}. In particular, the amplitude is determined by MHV-type building blocks, independently of the helicity configuration. In a first part of this section we show that Conjecture 1 implies that in MRK the four-dimensional scattering equations have a unique solution. The amplitude in MRK is then determined uniquely by this solution. This is very similar to MHV amplitudes, where the CHY formula has support only on a single solution of the scattering equations. In a second part, we show that this solution gives the correct amplitudes when the CHY formula is localised on it. This shows that Conjecture 1 implies the expected factorisation of amplitudes in MRK. This gives at the same time strong support to the validity of our conjecture.

\subsection{An exact solution of the scattering equations}
\label{sec-MRK-solution}

The aim of this section is to show that our conjecture implies that in MRK the four-dimensional scattering equations admit a unique solution (up to $GL(2,\mathbb{C})$ redundancy), independently of the number of external legs $n$ and the helicity configuration. In order to present the result, it is useful to introduce the following notation: $\overline{\mathfrak{N}}_{<a}$ ($\overline{\mathfrak{N}}_{>a}$) denotes the subset of $\overline{\mathfrak{N}}\equiv\mathfrak{N}\setminus\{1,2\}$ of elements that are less (greater) than $a$.

One of the main results of our paper is the following:
\begin{center}
\begin{mdframed}
If the incoming particle labels 1 and 2 are put into the set $\mathfrak{N}$, then the unique solution of the four-dimensional scattering equations  in MRK is, for $3\le a\le n-1$,
\begin{align}
 \label{eq:t_solution_MRK}t_a \,=\, \sqrt{k_a^+}&\times\left\{\begin{array}{ll}
\displaystyle {-1 \over \sqrt{\kappa}} \Bigg(\! \prod_{I\in\overline{\mathfrak{N}}_{<a}}\frac{q_{I}^{\perp}}{q_{I+1}^{\perp}}\!\Bigg)^*\,,&\textrm{ if $a\in\mathfrak{P}$,}\\
\displaystyle {\sqrt{\kappa} \over q_{a+1}^\perp}\Bigg(\!\prod_{I\in\overline{\mathfrak{N}}_{>a}}\frac{q_{I}^{\perp}}{q_{I+1}^{\perp}}\!\Bigg)\,,&\textrm{ otherwise,}
\end{array}\right.
\\
\label{eq:s_solution_MRK}\sigma_a \,=\, \frac{k_a^+}{k_a^\perp}&\times\left\{\begin{array}{ll}
\displaystyle
\Bigg(\!\prod_{I\in\overline{\mathfrak{N}}_{<a}}\frac{q_{I}^{\perp}}{q_{I+1}^{\perp}}\!\Bigg)^\ast
\Bigg(\!\prod_{I\in\overline{\mathfrak{N}}_{>a}}\frac{q_{I}^{\perp}}{q_{I+1}^{\perp}}\!\Bigg)\,,&\textrm{ if $a\in\mathfrak{P}$,}\\
\displaystyle {k_a^\perp \over q_{a+1}^{\perp}}\,\Bigg(\!\frac{q_a^\perp}{k_{a}^\perp}\!\Bigg)^\ast
\Bigg(\!\prod_{I\in\overline{\mathfrak{N}}_{<a}}\frac{q_{I}^{\perp}}{q_{I+1}^{\perp}}\!\Bigg)^\ast
\Bigg(\!\prod_{I\in\overline{\mathfrak{N}}_{>a}}\frac{q_{I}^{\perp}}{q_{I+1}^{\perp}}\!\Bigg)\,,&\textrm{ otherwise,}
\end{array}\right.
\end{align}
where the $GL(2,\mathbb{C})$ redundancy has been fixed according to eq.\,\eqref{gauge-fixing-4d-se}.
If helicity is not conserved in the impact factors, then the four-dimensional scattering equations have no solution in MRK.
\end{mdframed}
\end{center}
In eqs.~\eqref{eq:t_solution_MRK} and~\eqref{eq:s_solution_MRK} we interpret a product over an empty range as 1. Before we show how to derive eqs.~\eqref{eq:t_solution_MRK} and~\eqref{eq:s_solution_MRK} from Conjecture 1, let us make some comments about this solution. First, we immediately see that the solution in eqs.~\eqref{eq:t_solution_MRK} and~\eqref{eq:s_solution_MRK} has the same hierarchy as the rapidities of the produced particles, in agreement with Conjecture 1. Second, we find it remarkable that in MRK we can explicitly solve the scattering equations for arbitrary multiplicity $n$ and arbitrary $k$-sectors.
For each configuration of particle labels in sector $k$, we find a solution of the four-dimensional scattering equations. In Appendix~\ref{app:4dSE} we show that the solutions $\sigma_a$ of the four-dimensional scattering equations also solve the $D$-dimensional ones.
It is very rare that one can solve the scattering equations for arbitrary multiplicities, and so far this has only been done for the MHV sector, cf.~eq.\,\eqref{MHV-sol} or eq.\,\eqref{eq:MHV_solution} \cite{Fairlie:1972zz, Roberts:1972ggn, Fairlie:2008dg} (see also ref.~\cite{Weinzierl:2014vwa, Dolan:2014ega}) and for a very special two parameter family of kinematics \cite{Kalousios:2013eca}. We can recover the MHV solution of eq.\,\eqref{eq:MHV_solution} from eqs.~\eqref{eq:t_solution_MRK} and~\eqref{eq:s_solution_MRK} in the case where all produced particles have positive helicity, $\mathfrak{P}=\{3,\cdots,n\}$. Indeed, in that case we have $\overline{\mathfrak{N}}_{<a}=\overline{\mathfrak{N}}_{>a}=\emptyset$, and so the products in eqs.~\eqref{eq:t_solution_MRK} and~\eqref{eq:s_solution_MRK} do not contribute, and we immediately recover eq.\,\eqref{eq:MHV_solution}. Finally, we note that here we only consider the case where the strong ordering in rapidities is aligned with the cyclic ordering of the partial amplitude, which dominates in MRK. In Section~\ref{app:other_cyclic_orders} we repeat the analysis for other colour-orderings and we show that they are subleading in MRK.

In the remainder of this section we show how to derive eqs.~\eqref{eq:t_solution_MRK} and~\eqref{eq:s_solution_MRK} from Conjecture 1. 
Our starting point are the four-dimensional scattering equations in eq.\,\eqref{SE-4d-twistor}. We have seen that the hierarchy of the MHV solution becomes apparent after fixing the $GL(2,\mathbb{C})$ redundancy according to eq.\,\eqref{gauge-fixing-4d-se}, and in the following we assume that $(\sigma_1,\sigma_2,t_1,t_2)$ have been fixed to these values. Moreover, we always assume that $\{1,2\}\subseteq \mathfrak{N}$ and we have used eq.\,\eqref{SEs-12-momentum-conserv} to eliminate the equations associated to 1 and 2. We follow the convention that elements of $\mathfrak{P}$ and $\overline{\mathfrak{N}}$ are denoted by lower- and upper-case letters respectively, e.g., $i\in \mathfrak{P}$ and $I\in \overline{\mathfrak{N}}$. 

While the four-dimensional scattering equations in eq.\,\eqref{SE-4d-twistor} are naturally written in terms of spinor variables, MRK is most naturally defined in terms of lightcone coordinates. We therefore start by rewriting the four-dimensional scattering equations in terms of lightcone coordinates. We first perform a rescaling\footnote{An obvious advantage is that according to our conjecture for the $\tau_a$ ($a>2$) have a simple behaviour in QMRK, i.e.~$\tau_a={\cal O}(1)$.}
\begin{align}\label{t-resacling}
  t_i = \tau_i\,{\lambda_i^1 \over \sqrt{\kappa}} =\tau_i\,\sqrt{k_i^+ \over {\kappa}}\textrm{~~~~and~~~~} t_I= \tau_I\,{\sqrt{\kappa\,k_I^+}  \over k_I^{\bot}}\,.
\end{align}
After this rescaling, the  scattering equations become equivalent to $({\cal S}_i^{\alpha},\bar{\cal S}_I^{\dot \alpha})=0$, with
\begin{align}
\begin{aligned}
  {\cal S}_i^{1} \,&\equiv\,  {1 \over \lambda^1_i}\, {\cal E}^{1}_i,
  \qquad
  {\cal S}_i^{2} \,\equiv\,  {\lambda^1_i \over k_i^\perp}\, {\cal E}^{2}_i,
  \qquad  i \in \mathfrak{P},
  \\
  \bar{\cal S}_I^{\dot 1} \,&\equiv\,  \lambda^2_I\,\bar{\cal E}^{\dot 1}_I,
  \qquad
  \bar{\cal S}_I^{\dot 2} \,\equiv\,  \lambda^1_I\,\bar{\cal E}^{\dot 2}_I,
  \qquad  I \in \overline{\mathfrak{N}}\,,
  \\
  \bar{\cal S}_1^{\dot 1}  \,&\equiv\,  \lambda^2_1\,\bar{\cal E}^{\dot 1}_1, \qquad
  \bar{\cal S}_1^{\dot 2} \,\equiv\,  \lambda^2_1\,\bar{\cal E}^{\dot 2}_1, \qquad
  \bar{\cal S}_2^{\dot 1} \,\equiv\,  \lambda^1_2\,\bar{\cal E}^{\dot 1}_2, \qquad
  \bar{\cal S}_2^{\dot 2} \,\equiv\,  \lambda^1_2\,\bar{\cal E}^{\dot 2}_2\,.
\end{aligned}
\end{align}
In terms of lightcone coordinates, the new equations are explicitly given by
\begin{align}\label{4dSE-lightcone}
&\begin{aligned}
  {\cal S}_i^{1}  \,&=\,
  1  + \tau_i  - \sum_{I\in\overline{\mathfrak{N}}} {\tau_i \tau_I \over \sigma_i {-} \sigma_I} {k_I^+  \over  k_I^\perp}
  \,=\, 0,
  \\
  {\cal S}_i^{2}  \,&=\,
  1 + {k_i^+ \over k_i^\perp}{\tau_i \over \sigma_i}
  - {k_i^+ \over k_i^\perp} \sum_{I\in\overline{\mathfrak{N}}} {\tau_i \tau_I \over \sigma_i {-} \sigma_I}  
  \,=\, 0,
  \\
  \bar{\cal S}_1^{\dot 1} \,&=\, - \sum_{i\in\mathfrak{P}} {\tau_i \over \sigma_i} {k_i^+}
  \,=\, 0,
  \\
  \bar{\cal S}_1^{\dot 2} \,&=\, - {\kappa}  - \sum_{i\in\mathfrak{P}} {\tau_i \over \sigma_i} ({k_i^\perp})^\ast
  \,=\, 0,
\end{aligned}
\quad
\begin{aligned}
  \bar{\cal S}_I^{\dot 1}  \,&=\,
  {k_I^\perp}  - \sum_{i\in\mathfrak{P}} {\tau_i \tau_I \over \sigma_I {-} \sigma_i} {k_i^+}
  \,=\, 0,
  \\
  \bar{\cal S}_I^{\dot 2}  \,&=\,
  (k_I^\perp)^\ast  - {k_I^+ \over k_I^\perp} \sum_{i\in\mathfrak{P}} {\tau_i \tau_I \over \sigma_I {-} \sigma_i} ({k_i^\perp})^\ast
  \,=\, 0,
  \\
  \bar{\cal S}_2^{\dot 1}  \,&=\, -{\kappa}  - \sum_{i\in\mathfrak{P}} \tau_i {k_i^+}
  \,=\, 0,
  \\
  \bar{\cal S}_2^{\dot 2} \,&=\, - \sum_{i\in\mathfrak{P}} \tau_i ({k_i^\perp})^\ast
   \,=\, 0.
\end{aligned}
\end{align}
We stress that no Regge-limit has been applied to the previous equations, and they are fully equivalent to the four-dimensional scattering equations in eq.\,\eqref{SE-4d-twistor}, up to fixing the $GL(2,\mathbb{C})$ redundancy and performing the rescaling in eq.\,\eqref{t-resacling}.

Next we expand the scattering equations to leading power in MRK, taking into account that according to Conjecture 1 the solutions scale together with the lightcone +-components in the limit.
We write ${\cal S}_a^{\alpha} = S_a^{\alpha} + \ldots$ and $\overline{\cal S}_a^{\dot \alpha} = \overline{S}_a^{\dot\alpha} + \ldots$, where the dots indicate terms that are power-suppressed in the limit. We find
\beq\bsp
  \label{SE-MRK-p1} 
    { S}_i^{1}  =1 + \tau_i\,\Bigg(1+ \sum_{I\in\overline{\mathfrak N}_{<i}} \zeta_I \Bigg) = 0\,,\qquad 
  & {\overline S}_I^{\dot{1}}  = {k_I^\perp}  + \tau_I \sum_{i\in\mathfrak{P}_{<I}} \zeta_i\, {k_i^\perp}  = 0\,,
  \\
  { S}_i^{2}  = 1 + \zeta_i\,\Bigg(1 - \sum_{I\in\overline{\mathfrak N}_{>i}} \tau_I\Bigg)   = 0\,,\qquad
  & {\overline S}_I^{\dot{2}}  = ({k_I^\perp})^\ast  - \zeta_I\,\sum_{i\in\mathfrak{P}_{>I}} \tau_i ({k_i^\perp})^\ast  = 0\,,
\esp\eeq
where the sets $\mathfrak{P}_{>I}$ and $\mathfrak{P}_{<I}$ are defined in the obvious way, and we have defined the rescaled variables
\begin{align}\label{zeta-vars}
  \zeta_a \,&\equiv {k_a^+ \over k_a^\perp}\,{\tau_a \over \sigma_a}, \quad 3 \le a \le n\,.
\end{align}

Let us rewrite the scattering equations~\eqref{SE-MRK-p1} as:
\beq\bsp
  \label{SE-MRK-p1-x}
  { S}_i^{1}  \,=\,  1 + a_i\, \tau_i  =\, 0\,,  \qquad
  &\bar{ S}_I^{\dot 2}  \,=\,   ({k_I^\perp})^\ast  + b_I\,\zeta_I  =\, 0\,,
  \\
  { S}_i^{2} \,=\, 1 + c_i\, \zeta_i =\, 0\,,   \qquad  
  &\bar{ S}_I^{\dot 1}  \,=\,  {k_I^\perp}  + d_I\,\tau_I  =\, 0\,, 
\esp\eeq
with
\beq\bsp
  a_i \,\equiv\,  1 + \sum_{I\in\overline{\mathfrak N}_{<i}} \zeta_I\,,\qquad &
  b_I \,\equiv\, - \sum_{i\in\mathfrak{P}_{>I}} \tau_i\, ({k_i^\perp})^\ast,
  \\
  \label{SE-MRK-p2-x}
  c_i \,\equiv\, 1 -  \sum_{I\in\overline{\mathfrak N}_{>i}} \tau_I\,,  \qquad &
  d_I \,\equiv\,  \sum_{i\in\mathfrak{P}_{<I}} \zeta_i\, {k_i^\perp}.
\esp\eeq
We observe that in MRK the equations $({ S}_i^{1},{\bar S}_I^{\dot{2}})=0$ only depend on the variables $\tau_i$ and $\zeta_I$, while the equations $({ S}_i^{2},{\bar S}_I^{\dot{1}})=0$ only depend on $\tau_I$ and $\zeta_i$. We also see that each equation in eq.\,\eqref{SE-MRK-p1-x} is linear in one of the variables. We can use this very special structure to find an explicit analytic solution of the scattering equations in MRK.

To start, we can use the fact that the equations ${S}_i^\alpha=0$ are linear in $\tau_i$ and $\zeta_i$ to obtain
\begin{align}\label{ti-zi-formal}
  \tau_i \,=\, - {1 \over a_i},  \qquad
  \zeta_i \,=\, - {1 \over c_i}.
\end{align}
Inserting eq.\,\eqref{ti-zi-formal} into the expressions for $b_I$ and $d_I$ in eq.\,\eqref{SE-MRK-p2-x}, we find
\beq\bsp
  \label{b-coe}
  b_I \,=\, \sum_{i\in\mathfrak{P}_{>I}} {({k_i^\perp})^\ast \over a_i}
  \,&=\, \sum_{i\in\mathfrak{P}_{>I}} {({k_i^\perp})^\ast}\,\Bigg(1 + \sum\limits_{J\in\overline{\mathfrak N}_{<i}} \zeta_J\Bigg)^{\!\!-1}\,,
  \\
  %
  d_I \,=\,  -\sum_{i\in\mathfrak{P}_{<I}} {k_i^\perp \over c_i}
  \,&=\, - \sum_{i\in\mathfrak{P}_{<I}} {k_i^\perp}\,\Bigg(1 -  \sum\limits_{J\in\overline{\mathfrak N}_{>i}} \tau_J\Bigg)^{\!\!-1}\,.
\esp\eeq
The $b_I$ are solely determined by the variables $\zeta_I$, and similarly the variables $d_I$ are determined by the $\tau_I$.
In order to proceed, we write $\overline{\mathfrak N}=\{I_\ell\}_{1\le \ell\le m}$ ($m\equiv k-2$) with $I_{\ell}<I_{\ell+1}$. The quantities $b_I$ and $d_I$ satisfy the recursions, for $1 \le r \le m$,
\begin{align}\label{b-recursion}
  b_{I_r}   \,&=\, b_{I_{r+1}}  +  \Bigg(1 + \sum_{l=1}^{r} \zeta_{I_l}\Bigg)^{\!\!-1}\,(q_{I_{r+1}}^\perp - q_{I_{r}+1}^{\perp})^\ast\,,\\
  d_{I_r}   \,&=\, d_{I_{r-1}}  -  \Bigg(1 - \sum_{l=r}^{m} \tau_{I_l}\Bigg)^{\!\!-1}\,(q_{I_{r}}^\perp - q_{I_{r-1}+1}^{\perp})\,.
\end{align}
The recursion starts with $b_{I_{m+1}} = d_{I_0} = q_{I_{m+1}} = q_{I_0+1}=0$.
Indeed, we have
\begin{align}
\begin{aligned}
  b_{I_r} \,&=
  \sum_{i\in\mathfrak{P}_{>I_r}} ({{k_i^\perp})^\ast}\,\Bigg(1 + \sum\limits_{J\in\overline{\mathfrak N}_{<i}} \zeta_J\Bigg)^{\!\!-1}
  \\
  \,&=
  \Bigg( \sum_{i\in\mathfrak{P}_{>I_{r+1}}} + \sum_{I_r < i < I_{r+1}} \Bigg) (k_i^\perp)^\ast
  \,\Bigg(1 + \sum\limits_{J\in\overline{\mathfrak N}_{<i}} \zeta_J\Bigg)^{\!\!-1}
  \\
  \,&= b_{I_{r+1}}  +  \Bigg(1 + \sum\limits_{l=1}^{r} \zeta_{I_l}\Bigg)^{\!\!-1}
  \sum_{\substack{I_r < i < I_{r+1}}}({k_i^\perp})^\ast\\
  \,&=b_{I_{r+1}}  +  \Bigg(1 + \sum_{l=1}^{r} \zeta_{I_l}\Bigg)^{\!\!-1}\,(q_{I_{r+1}}^\perp - q_{I_{r}+1}^{\perp})^\ast\,.
\end{aligned}
\end{align}
The proof of the recursion for $d_{I_r}$ is similar. We can combine the recursions for $b_I$ and $d_I$ with the scattering equations and turn them into recursions for $\tau_I$ and $\zeta_I$. We illustrate this procedure explicitly on $b_I$ and $\tau_I$. The procedure for $d_I$ and $\zeta_I$ is similar. We start from the equation $\bar{ S}_{I_r}^{\dot 1}=0$ and insert the recursion in eq.\,\eqref{b-recursion}. For $r=1$, we find
\begin{align}
\label{eq:S_I_1^1}
  0=\bar{ S}_{I_1}^{\dot 1}  \,=\,  {k_{I_1}^\perp}  + d_{I_1} \tau_{I_1}  \,=\,
  - \Bigg(1 -  \sum\limits_{l=1}^m \tau_{I_l}\Bigg)^{\!\!-1}
  \left[ -{k_{I_1}^\perp}\left(1 - \sum_{l=2}^m \tau_{I_l} \right) + \tau_{I_1} q_{I_1+1}^{\perp} \right]
  \,.
\end{align}
This immediately leads to
\begin{align}\label{t-I-recursion-1}
  \tau_{I_1}  \,=\,  {k_{I_1}^\perp  \over q_{I_1+1}^\perp} \left( 1 - \sum_{l=2}^m \tau_{I_l} \right).
\end{align}
Similarly, using eq.\,\eqref{t-I-recursion-1} we obtain for $r=2$,
\beq\bsp
 0&\,= \bar{ S}_{I_2}^{\dot 1}  =\,   {k_{I_2}^\perp}  + d_{I_2} \tau_{I_2}  \\
 &\, =  {k_{I_2}^\perp}  - \tau_{I_2}
  \Bigg[ {k_{I_1}^\perp \over \tau_{I_1}} + \Bigg(  1 -  \sum\limits_{l=2}^m \tau_{I_l}\Bigg)^{\!\!-1} (q^{\perp}_{I_2} - q^{\perp}_{I_1+1}) \Bigg]\\
  &\, = \Bigg(1 -  \sum\limits_{l=2}^m \tau_{I_l}\Bigg)^{\!\!-1}
  \left[  {k_{I_2}^\perp}\left( 1 - \sum_{l=3}^m \tau_{I_l} \right) - \tau_{I_2} q_{I_2+1}^{\perp} \right]
  \,,
\esp\eeq
and so we have
\begin{align}\label{t-I-recursion-2}
  \tau_{I_2}  \,=\, {k_{I_2}^\perp  \over q_{I_2+1}^\perp} \left( 1 - \sum_{l=3}^m \tau_{I_l} \right)\,.
\end{align}
A similar formula holds for general $r$, and the $\tau_{I_r}$ satisfy the recursion,
\begin{align}\label{t-I-recursion-r}
  \tau_{I_r}  \,=\,  {k_{I_r}^\perp  \over q_{I_r+1}^\perp} \left( 1 - \sum_{l=r+1}^m \tau_{I_l} \right)\,.
\end{align}
The recursion starts with 
\begin{align}\label{t-I-m}
  \tau_{I_m}  \,=\,  {k_{I_m}^\perp  \over q_{I_m+1}^\perp}\,.
\end{align}
It is easy to show by induction that the recursion admits the explicit solution
\begin{align}\label{t-I-r}
  \tau_{I_r}  \,=\, {k_{I_r}^\perp  \over q_{I_r+1}^\perp}\, \prod_{l=r+1}^{m} {q_{I_l}^\perp  \over q_{I_l+1}^\perp}\,, \qquad 1 \le r < m\,.
\end{align}

The explicit solution for the variables $\zeta_{I_r}$ can be obtained in the same way. More precisely, one can show that there is a recursion
\begin{align}
\begin{aligned}
  \zeta_{I_r}  \,&=\, \Bigg({k_{I_r}^{\perp}  \over q_{I_r}^{\perp}}\Bigg)^{\!\!\ast}\, \left(1+ \prod_{l=1}^{r-1} \zeta_{I_l}\right)\,, \qquad 1 < r \le m\,,
\end{aligned}
\end{align}
with
\beq\bsp
  \zeta_{I_1}  =\,  \Bigg({k_{I_1}^{\perp} \over q_{I_1}^{\perp}}\Bigg)^{\!\!\ast}\,,
  \esp\eeq
  and the recursion admits the explicit solution
  \begin{align}\label{z-I-final}
\begin{aligned}
  \zeta_{I_r}  \,&=\, \Bigg( {k_{I_r}^{\perp}  \over q_{I_r}^{\perp}}\Bigg)^{\!\!\ast}\, \Bigg(\prod_{l=1}^{r-1} {q_{I_l+1}^{\perp} \over q_{I_l}^{\perp}}\Bigg)^{\!\!\ast}, \qquad 1 < r \le m\,.
\end{aligned}
\end{align}

Together with eq.\,\eqref{ti-zi-formal}, eqs.~\eqref{t-I-r} and~\eqref{z-I-final} provide the explicit solution of the scattering equations in MRK. We see that there is indeed a unique (independent) solution to the four-dimensional scattering equations in MRK, which can be traced back to the fact that at every step we only needed to solve linear equations.
We can easily obtain explicit solutions for the coefficients $c_i$ and $a_i$ that appear in eq.\,\eqref{ti-zi-formal}. For example, using eq.\,\eqref{t-I-recursion-r} and~\eqref{t-I-r}, we find
\begin{align}
  \label{c_i}
  c_i \,=\, 1 -  \sum_{I\in\overline{\mathfrak{N}}_{>i}} \tau_{I} \,=\,  \prod_{I\in\overline{\mathfrak{N}}_{>i}} {q_{I}^\perp  \over q_{I+1}^\perp}\,.
  \end{align}
  We recall that we use the convention that products over empty ranges are unity.
Similarly for $a_i$, we have
\begin{align}
  \label{a_i}
  a_i  &=  1 + \sum_{I\in\overline{\mathfrak{N}}_{<i}} \zeta_{I} 
  \,=\, \Bigg(\prod_{I\in\overline{\mathfrak{N}}_{<i}} {q_{I+1}^{\perp} \over q_{I}^{\perp}}\Bigg)^{\!\!\ast}\,.
\end{align}
Hence, we find
\beq
\tau_i \,=\, - \Bigg(\prod_{I\in\overline{\mathfrak{N}}_{<i}} {q_{I}^{\perp} \over q_{I+1}^{\perp}}\Bigg)^{\!\!\ast}
\text{~~~~and~~~~} 
\zeta_i \,=\,  -\prod_{I\in\overline{\mathfrak{N}}_{>i}} {q_{I+1}^\perp  \over q_{I}^\perp}\,.
\eeq
Finally, we recover eqs.\,\eqref{eq:t_solution_MRK} and~\eqref{eq:s_solution_MRK} by changing variables from $(\tau_a,\zeta_a)$ to $(t_a,\sigma_a)$ using eqs.\,\eqref{t-resacling} and~\eqref{zeta-vars}.


\subsection{Multi-Regge factorisation from the CHY equation}
In the previous section we have shown that Conjecture 1 implies that in MRK the scattering equations have a unique solution (up to $GL(2,\mathbb{C})$ redundancy). In this section we show that Conjecture 1 also implies that any $n$-point gluon amplitude in MRK can be written in the factorised form in eq.\,\eqref{amp-MRK-main}. Since eqs.\,\eqref{eq:t_solution_MRK} and~\eqref{eq:s_solution_MRK} hold for arbitrary helicity configurations and multiplicities, this shows that MRK factorisation holds independently of the helicity configuration and the number of legs. This nicely complements the results of ref.~\cite{DelDuca:1995zy}, where factorisation was shown to hold for the simplest helicity configurations. We emphasise, however, that we cannot present at this point a rigorous proof that multi-Regge factorisation holds for arbitrary helicity configurations, because our derivation relies on Conjecture 1. Finding a rigorous mathematical proof of Conjecture 1 would thus eventually imply an elegant proof of the factorisation of tree-level gluon amplitudes in MRK, for arbitrary numbers of external legs and helicity configurations.

We start by considering $n$-gluon scattering in the case where the two incoming gluons (labeled by $1$ and $2$ respectively) carry the same helicity, and we comment on the case where they have different helicities at the end of this section.
If we fix the $GL(2,\mathbb{C})$ redundancy in eq.\,\eqref{amp-YM-4d} according to eq.\,\eqref{gauge-fixing-4d-se} and use four $\delta$-functions and write them in terms of momentum conservation, then the formula for the amplitude reduces to
\begin{align}\label{amp-k-sector}
  {\cal A}&_n(1^-,2^-,  \ldots, n)
  \nonumber\\
  \,&=\, -s\, \bigintsss
  \prod\limits_{a=3}^{n} {d\sigma_a d\tau_a \over \tau_a}\,
  {1 \over \sigma_{34}\cdots\sigma_{n-1, n}\sigma_{n}}\,
  \left( \prod_{i\in\mathfrak{P}, I\in\overline{\mathfrak N}} {k_I^\perp \over k_i^\perp}  \right)
  \prod_{I\in\overline{\mathfrak N}} \delta^2 \big( \bar{\cal S}^{\dot\alpha}_I \big)
  \prod_{i\in\mathfrak{P}} \delta^2\big( {\cal S}^\alpha_i \big)\,.
\end{align}
We would like to emphasise that eq.~\eqref{amp-k-sector} is exact and no Regge limit has been applied to it.

From Conjecture 1 it follows that in MRK the solutions to the scattering equations satisfy $\sigma_{a,a+1} \simeq \sigma_a$. Inserting this relation into eq.\,\eqref{amp-k-sector} and passing to the rescaled variables defined in eqs.\,\eqref{t-resacling} and \eqref{zeta-vars}, we see that the formula for the amplitude in MRK can be written as
\begin{align}\label{amp-CHY-MRK-formal}
  {\cal A}_n&(1^-,2^-,  \ldots, n)
  \,\simeq\, -s\, \left(\bigintsss \prod\limits_{a=3}^{n} {d\tau_a d\zeta_a \over \zeta_a \tau_a} \right)\,
  \left( \prod_{i\in\mathfrak{P}} {1 \over k_i^\perp}\, \delta^2\big({ S}^\alpha_i\big) \right)
  \left( \prod_{I\in\overline{\mathfrak N}} k_I^\perp\, \delta^2\big(\bar{ S}^{\dot\alpha}_I\big)  \right)\,,
\end{align}
where the arguments of the $\delta$-functions are given in eq.\,\eqref{SE-MRK-p1}. We know that these equations have a unique solution given by eqs.~\eqref{eq:t_solution_MRK} and~\eqref{eq:s_solution_MRK} in eq.~\eqref{amp-MRK-main}. We now show that when the integral in eq.\,\eqref{amp-CHY-MRK-formal} is localised on this solution, then we recover the conjectured factorised form of a tree-level amplitude in MRK. We do this by a procedure very similar to the one used in the previous section to obtain the solutions in eqs.~\eqref{eq:t_solution_MRK} and~\eqref{eq:s_solution_MRK}. The proof of eqs.~\eqref{eq:t_solution_MRK} and~\eqref{eq:s_solution_MRK} relies heavily on the fact that in MRK the scattering equations decouple into a set of linear equations that can be solved one-by-one. Here we use this fact to solve the $\delta$-functions in the integrand of eq.\,\eqref{amp-CHY-MRK-formal} one at the time. All the manipulations are identical to those performed in the previous section, so we will be brief on the derivation and only highlight some aspects related to solving the $\delta$-functions.

We start by performing the integrations over  the $\tau_i$ and $\zeta_i$. From eq.\,\eqref{SE-MRK-p1} we see that these integrations are independent from each other. We can localise them completely using $\delta\big({ S}_i^{1}\big)$ and $\delta\big({ S}_i^{2}\big)$, and we find
\beq\bsp\label{}
  \int {d\tau_i \over \tau_i}\, \delta\big({ S}_i^{1}\big)  
  \,&=\,  \int {d\tau_i \over \tau_i}\, \delta\big(1 + a_i \tau_i\big)
  \,=\, -1\,,
  \\
  \int {d\zeta_i \over \zeta_i}\, \delta\big({ S}_i^{2}\big)  
  \,&=\,  \int {d\zeta_i \over \zeta_i}\, \delta\big(1 + c_i \zeta_i \big)
  \,=\, - 1\,.
\esp\eeq
After this step all the variables $\tau_i$ and $\zeta_i$ have been integrated out.

Using the same reasoning as in the previous section, we see that the integrations over $\tau_I$ and $\zeta_I$ are now independent from each other, and so we can discuss the integrations over the $\tau_I$ variables independently from the integrations over the $\zeta_I$ variables. 
Let us start with the $\tau_I$-integrations. We use  the $\delta$-functions $\delta\big(\bar{ S}^{\dot 1}_I\big)$ combined with the recursive procedure of the previous section to localise all the $\tau_I$ variables. In particular, we localise the $\tau_I$ variables in increasing order from $I_1$ to $I_m$. 
Let us look at the integration over $\tau_{I_1}$. We start from eq.\,\eqref{eq:S_I_1^1} to obtain
\beq\bsp
  \int {d\tau_{I_1}  \over  \tau_{I_1}}\, \delta\big(\bar{ S}^{\dot 1}_{I_1}\big)
&  \,=\, \oint_{\cal C} {d\tau_{I_1}  \over  \tau_{I_1}}\, {1 \over  \bar{ S}^{\dot 1}_{I_1}}
  \,=\, \oint_{\cal C} {d\tau_{I_1}  \over  \tau_{I_1}}
   {1- \sum_{l=1}^m \tau_{I_l}    \over
   {k_{I_1}^\perp} \big(1 - \sum_{l=2}^m \tau_{I_l}\big) - \tau_{I_1} q_{I_1+1}^\perp}\\
  &  \,=\, - {1 \over k_{I_1}^\perp}\, {q_{I_1}^\perp   \over q_{I_1+1}^\perp}\,,
\esp\eeq
where we have interpreted the integral over the $\delta$-function as a contour integral over a contour $\cC$ encircling the zero of the argument of the $\delta$-function. 
The remaining integrals over the $\tau_I$ variables can be performed in the same way {in increasing order}, with the simple result
\begin{align}
  \int {d\tau_{I}  \over  \tau_{I}}\, \delta\big(\bar{ S}^{\dot 1}_{I}\big)
  \,=\,  - {1 \over k_{I}^\perp}\, {q_{I}^\perp   \over q_{I+1}^\perp}\,.
\end{align}
Finally, we are left with the integrals over the $\zeta_I$ variables to perform. The procedure is the same as for the $\tau_I$ variables, combined with the recursive procedure in the previous section. We find
\begin{align}
  \int {d\zeta_{I}  \over \zeta_{I}}\, \delta\big(\bar{ S}^{\dot 2}_{I}\big)
  \,=\,  - \Bigg({1 \over k_I^{\perp}}\, {q_{I+1}^\perp   \over q_{I}^\perp}\Bigg)^{\!\! \ast}.
\end{align}

We see that we can perform all integrations one after the other. As a result, the amplitude takes a completely factorised form, which has its origin in the fact that in MRK all the integrations in the CHY formula decouple and can be performed one-by-one. The final result takes the very simple factorised form in eq.\,\eqref{amp-MRK-main}. Conjecture 1 thus implies the factorised form of the tree-level amplitude in MRK. Unlike previous derivations, we stress that our derivation is completely independent of the helicities of the produced particles, giving strong support to the idea that MRK-factorisation of tree-level amplitudes holds for any multiplicity and for any helicity assignment.

Finally, let us comment on what happens in the case where the incoming gluons $1$ and $2$ have opposite helicities, for example when gluon $1$ has positive helicity. Since helicity is conserved by the impact factor $C(1;n)$, we obtain a non-zero result in MRK only when gluon $n$ has negative helicity. One can show that the amplitudes $\cA_n(1^-,2^-,\ldots,n^+)$ and $\cA_n(1^+,2^-,\ldots,n^-)$ can be computed using the same CHY formula~\eqref{amp-YM-4d}, up to performing the following replacement on the integrand in eq.~\eqref{eq:YM_integrand_4d},
\beq\label{eq:I_YM_modif}
\cI_n(1^-,2^-,\ldots,n^+) = \frac{1}{(12)\ldots(n1)} \,\longrightarrow\, \frac{1}{(1n)^4}\,\cI_n(1^-,2^-,\ldots,n^+)\,.
\eeq
If the $GL(2,\mathbb{C})$ redundancy is fixed according to eq.\,\eqref{gauge-fixing-4d-se}, then the additional factor in the left-hand side of eq.\,\eqref{eq:I_YM_modif} only depends on the variables $t_n$ and $\sigma_n$. More precisely, we have,
\beq\label{eq:insert_eq}
  {1 \over (1\,n)} \,=\, {t_n \over \sigma_n} 
  \,=\, \sqrt{k_n^+ \over \kappa}\, {k_n^\perp \over k_n^+}\,\zeta_n\,.
\eeq
In MRK $\zeta_n$ is uniquely fixed by the equation $S_n^2 = 1 + \zeta_n=0$. Thus we have
\beq
  \left.{1 \over (1n)^4}\right|_{\zeta_n=-1} \,=\, \bigg( {k_n^\perp \over (k_n^\perp)^\ast} \bigg)^2\,.
\eeq
This phase factor combines with the impact factor $C(1^-;n^+)$ to give
\beq
C(1^-;n^+) \, \bigg( {k_n^\perp \over (k_n^\perp)^\ast} \bigg)^2 =  {k_n^\perp \over (k_n^\perp)^\ast}  = C(1^+;n^-)\,.
\eeq
We see that in the case where $h_1=-h_2=+1$, only the impact factor $C(1;n)$ changes, in agreement with the factorisation of tree-level amplitudes in MRK.


\subsection{Amplitudes where the cyclic colour ordering is not aligned with the rapidity ordering}
\label{app:other_cyclic_orders}

In the previous section, we have shown how the MRK factorisation of tree-level amplitudes where the cyclic colour-ordering is aligned with the rapidity ordering follows from the CHY representation of the amplitude and Conjecture 1. Here we show that all other cyclic colour orderings are suppressed in MRK. This complements nicely the results of ref.~\cite{DelDuca:1995zy}, where this result had been shown to hold only for MHV amplitudes.

We have seen that in MRK the scattering equations have a unique solution. As a consequence, we can obtain a very simple relation between amplitudes in MRK with different cyclic orderings,
\begin{align}\label{Amp-non-canonical-ordering}
  {\cal A}^{\textrm{MRK}}_n(1^-,2^-, \mu_3,  \ldots, \mu_n)
  \,=\, {\cal A}^{\textrm{MRK}}_n(1^-,2^-, 3,  \ldots, n)\, {\cal R}[\mu]\,,
\end{align}
where ${\cal A}^{\textrm{MRK}}_n$ is the $n$-point amplitude in MRK and $(\mu_3,\ldots,\mu_n)$ is a permutation of $(3,\ldots,n)$ and we defined,
\begin{align}\label{}
  {\cal R}[\mu] \,=\,  
  {\sigma_3 \sigma_4 \cdots \sigma_n \over \sigma_{\mu_3 \mu_4}\sigma_{\mu_4 \mu_5} \cdots \sigma_{\mu_{n-1} \mu_n} \sigma_{\mu_n}} 
  \bigg|_{\text{MRK solution \eqref{eq:s_solution_MRK}}}
  \,\equiv\,  \prod_{i=3}^{n-1} {\cal R}_{\mu_i} \bigg|_{\text{MRK solution \eqref{eq:s_solution_MRK}}}\,,
\end{align}
with
\beq
{\cal R}_{\mu_i} \,\equiv\, {\sigma_{\mu_i} \over \sigma_{\mu_i \mu_{i+1}}}\,.
\eeq
Our conjecture implies that
\begin{align}\label{}
  {\cal R}_{\mu_i} \,
  \,\simeq\, 
  \left\{\begin{array}{ll}
        1, & \mu_i < \mu_{i+1}  \\
        - {\sigma_{\mu_i} \over \sigma_{\mu_{i+1}}},  & \mu_i > \mu_{i+1}
  \end{array}\right.
  \,=\,
  \left\{\begin{array}{ll}
        1, & \mu_i < \mu_{i+1}\,,  \\
        {\cal O}\big(k_{\mu_i}^+ / k_{\mu_{i+1}}^+ \big),  & \mu_i > \mu_{i+1}\,.
  \end{array}\right.
\end{align}
It is then easy to see that ${\cal R}[\mu]$ is suppressed in MRK kinematics unless the cyclic colour-ordering is aligned with the rapidity ordering, in agreement with known results for MHV amplitudes~\cite{DelDuca:1995zy}.


\section{Quasi-Multi-Regge factorisation from the scattering equations}
\label{sec-QMRK}

In the previous section we have shown that one can derive the factorisation of tree-level gluon amplitudes in MRK from the CHY formula and Conjecture 1. One feature of our derivation is that it is independent of the number $n$ of external gluons as well as their helicities. In this section we extend this analysis to various quasi-multi-Regge limits. More precisely, we investigate quasi-multi-Regge limits where the produced particles are strongly ordered in rapidity, except for a single cluster of particles for which no rapidity-ordering is imposed.
We will show that, in agreement with expectations from Regge theory, in each of these two cases Conjecture 1 implies that the amplitude factorises into a set of universal building blocks which are multi-particle generalisations of the impact factors and Lipatov vertices in the case of MRK, independently of the multiplicity and the helicities of the produced particles. We also obtain CHY-type representations for these building blocks.

\subsection{The quasi-multi-Regge limit $y_3 \simeq\cdots\simeq y_r \gg \cdots \gg y_n$}
\label{sec:qmrk_C}
The goal of this section is to show that Conjecture 1 implies that in the quasi-multi-Regge limit where $y_3 \simeq\cdots\simeq y_r \gg \cdots \gg y_n$ any tree-level gluon amplitude factorises as follows:
\begin{align}\label{eq:Amp_C_QMRK}
  {\cal A}_n&(1,\!\ldots\!,n)
  \\
  \nonumber
  \simeq&\, s\,C(2;3,\ldots,r)\, {-1 \over |{q_{r+1}^\perp}|^2}\, V(q_{r+1}; r{+}1; q_{r+2}) \cdots
  {-1 \over |{q_{n-1}^\perp}|^2}\, V(q_{n-1}; n{-}1; q_{n})\,
  {-1 \over |{q_{n}^\perp}|^2}\, C(1; n)\,,
\end{align}
where $C(2;3,\ldots,r)$ is a generalised impact factor that only depends on the subset of momenta $(k_2, \ldots, k_r)$. We will now show that the generalised impact factors are universal, i.e., they do not depend on the quantum numbers of the other particles involved in the scattering.

Let us start by analysing the case $r=n{-}1$. We assume without loss of generality that particles $1$ and $n$ have negative and positive helicities respectively (if they have the same helicity, the scattering equations have no solution and the amplitude vanishes -- see Appendix~\ref{app:conjecture}--, in agreement with the fact that helicity is conserved in the impact factor $C(1;n)$). For simplicity we also assume that $h_2=-1$. The case $h_2=+1$ can be recovered by using the same argument at the end of the previous section. The general logic will be similar to the MRK case, so we will be brief and we will not describe all the steps in detail here. We start by fixing the $GL(2,\mathbb{C})$ redundancy as in eq.\,\eqref{gauge-fixing-4d-se}, and we apply Conjecture 1 to expand the scattering equations to leading order in the limit. We write $(\mathcal{S}_i^{\alpha},\overline{\mathcal{S}}_I^{\dot\alpha}) = ({S}_i^{\alpha},\overline{{S}}_I^{\dot\alpha}) + \cdots$, where the dots indicate terms that are power-suppressed in the limit. We observe that there is a subset of scattering equations that become linear {in $\tau_n$ and $\zeta_n$},
  \beq\bsp
  \label{4dSE-QMRK-n-n1}
  {S}_n^{ 1}  \,&=\, 1 +\tau_n \Bigg(1 +  \sum_{I\in\overline{\mathfrak N}} \zeta_I \Bigg)
  \,=\, 0\,,
\\
  {S}_n^{ 2}  \,&=\, 1 + \zeta_n
  \,=\, 0\,,
\esp\eeq
where we use the rescaled variables defined in eqs.\,\eqref{t-resacling} and~\eqref{zeta-vars}. We can easily evaluate the corresponding residues,
\beq\bsp
  {({q_n^\perp})^\ast \over k_n^\perp}\, \int {d\tau_n \over \tau_n} \int {d\zeta_n \over \zeta_n}\,
  \delta\big({S}^{1}_n\big) \delta\big({S}^{2}_n\big)
  \,=\,   {({k_n^\perp})^\ast \over {k_n^\perp}} =   C(1^-; n^+) \,,
  \esp\eeq
  where we used the fact that $q_n^\perp = - k_n^\perp$. The amplitude then takes the expected factorised form
  \beq
  \!\!\!{\cal A}_n(1^-,2^-,\!\ldots\!,n^+) \,\simeq\, s\, C(2^-;3,\ldots,n{-}1)\, {-1 \over |{q_{n}^\perp}|^2}\, C(1^-; n^+)\,,
\eeq
where $C(1^-; n^+)$ is the same impact factor as in MRK, cf.~eq.\,\eqref{eq:impact_factors}. The generalised impact factor $C(2^-;3,\ldots,n{-}1)$ admits a CHY-type representation,
\begin{align}
  C\big(2^-; 3&, \ldots, n{-}1\big)
  \nonumber
  \\
  \,=\, {q_n^\perp}&\bigintsss
  \prod\limits_{a=3}^{n-1} {d\sigma_a d\tau_a \over \tau_a}\,
  {1 \over \sigma_{34}\cdots\sigma_{n-2, n-1}\sigma_{n-1}}\,
  \left( \prod_{i\in\mathfrak{P}, I\in\overline{\mathfrak N}} {k_I^\perp \over k_i^\perp}  \right)
  \nonumber
  \\
  \times&\prod_{I\in\overline{\mathfrak N}} 
  \delta\Bigg( {k_I^\perp}  - \sum_{i\in\mathfrak{P}} {\tau_I \tau_i \over \sigma_I {-} \sigma_i} {k_i^+} \Bigg)
  \delta\Bigg( (k_I^\perp)^\ast  - {k_I^+ \over k_I^\perp} \sum_{i\in\mathfrak{P}} {\tau_I \tau_i \over \sigma_I {-} \sigma_i} ({k_i^\perp})^\ast
  - \zeta_I\, {({q_n^\perp})^\ast \over  1 + \sum_{J\in\overline{\mathfrak N}} \zeta_J} \Bigg)
  \nonumber
  \\
  \times&\prod_{i\in\mathfrak{P}}
  \delta\Bigg( 1 + \tau_i  - \sum_{I\in\overline{\mathfrak N}} {\tau_i \tau_I \over \sigma_i {-} \sigma_I} {k_I^+  \over  k_I^\perp}   \Bigg)
  \delta\Bigg( 1 + \zeta_i  - {k_i^+ \over k_i^\perp}\sum_{I\in\overline{\mathfrak N}} {\tau_i \tau_I \over \sigma_i {-} \sigma_I} \Bigg),
  \label{Impact-factor-main}
\end{align}
where $\overline{\mathfrak N}={\mathfrak N}\backslash\{2\}$ and $q_n=\sum_{a=2}^{n-1}k_a $ is the total momentum exchanged in the $t$-channel. A similar formula can be derived when $h_2= {+}1$. We have checked that our formula correctly reproduces various results for impact factors in the literature~\cite{DelDuca:1999iql,DelDuca:1995ki,Duhr:2009uxa}. In particular, eq.\,\eqref{Impact-factor-main} correctly reproduces known results for the MHV-type impact factors $C\big(2^-; 3^+, \ldots, (n{-}1)^+\big)$ for arbitrary multiplicities. 
Moreover, we have checked that we reproduce the NMHV results for up to $n=7$. In Appendix~\ref{app:factorisations} we also show the the formula in eq.\,\eqref{Impact-factor-main} has the factorisation properties expected from a gluon amplitude at tree level. In particular, if gluon $i$ becomes soft, $k_i\to0$, the impact factor factorises into an impact factor with one particle less, times the usual eikonal factor \cite{Berends:1988zn}.
{For example, if the soft gluon carries positive helicity, we have}
\beq
  C\big(2; 3, \ldots, i,\ldots, n{-}1\big)  \,\simeq\,   C\big(2; 3, \ldots, a,b, \ldots, n{-}1\big) \,
  {\braket{a\,b} \over \braket{a\,i} \braket{i\,b}}\,,
\eeq
where $(a,b) = (i{-}1,i{+}1)$ are the gluons adjacent to $i$ for the chosen colour-ordering. Similarly, if two produced gluons, say $i$ and $i{+}1$, become collinear, we have
\beq
  C\big(2; 3, \ldots, n{-}1\big)  \,\simeq\,  C\big(2; 3, \ldots, i{-}1, K^h\!, i{+}2, \ldots, n{-}1\big) \operatorname{Split}_{-h}\!\big(i,i{+}1\big)\,,
\eeq  
where $K = k_i+k_{i+1}$ denotes the momentum of the parent gluon before the splitting, and Split denotes the usual tree-level splitting function. Finally, in the case where particle $n{-}1$ has much smaller rapidity as the other produced gluons, we have
\beq\label{eq:C_r_V}
  C\big(2; 3, \ldots, n{-}1\big) \,\simeq\, C\big(2; 3, \ldots, n{-}2\big) \,{-1 \over |q_{n-1}^{\perp}|^2}\,V(q_{n-1}; n{-}1; q_n)\,,
\eeq  
with $q_n = q_{n-1} + k_{n-1} = k_2 + k_{3} + \cdots +k_{n-1}$.

The previous considerations also imply that the amplitude has the expected factorisation behaviour in the quasi-multi-Regge limit where $y_3 \simeq\cdots\simeq y_r \gg \cdots \gg y_n$. Indeed, we can simply apply eq.\,\eqref{eq:C_r_V} iteratively to add one large rapidity gap at the time, and we immediately see that the amplitude takes the factorised form given in eq.\,\eqref{eq:Amp_C_QMRK}. We note that this factorisation is a direct consequence of Conjecture 1 and the CHY representation of gluon amplitudes, and it does not rely on any other assumptions. In particular, we see that the factorised form holds for arbitrary multiplicities and helicity configurations. Moreover, we see that the impact factors are universal, in the sense they do not depend on the quantum numbers of the other produced particles, in agreement with general expectations. This follows immediately from the existence of the CHY-type representation of the generalised impact factors in eq.\,\eqref{Impact-factor-main}.

\subsection{The quasi-multi-Regge limit $y_3\gg\cdots\gg y_r  \simeq\cdots\simeq y_s\gg  \cdots \gg y_n$}
In this section we extend our analysis to the quasi-multi-Regge limit $y_3\gg\cdots\gg y_r  \simeq\cdots\simeq y_s\gg  \cdots \gg y_n$ where the amplitude is expected to factorise as follows:
\begin{align}\label{eq:Amp_V_QMRK}
  {\cal A}&_n(1,\ldots,n) \\
  \nonumber\,&\simeq\, s\,C(2;3) {-1 \over |{q_{4}^\perp}|^2} V(q_{4}; 4; q_{5}) \cdots
 V(q_{r}; r,\ldots,s; q_{s+1}) \cdots
 V(q_{n-1}; n{-}1; q_{n}) {-1 \over |{q_{n}^\perp}|^2} C(1; n)\,.
\end{align}
We start by analysing the case $(r,s) = (4,n{-}1)$, and we assume that $(h_1,h_2)=(-1,-1)$. Helicity conservation implies that we obtain a non-zero result only if $(h_3,h_n)=(+1,+1)$, which we assume from now on.
We proceed in the usual way, and we use Conjecture 1 to expand the scattering equations to leading order in the limit. We then observe that the equations $S_3^{\alpha} = S_n^{\alpha}=0$ are linear in $\zeta_i$ and $\tau_i$, $i\in\{3,n\}$. We can thus localise the integrals over these variables on the residues obtained by solving these linear equations. We arrive at the following factorised form for the amplitude:
\begin{align}
  {\cal A}_n(1^-,2^-,3,\!\ldots\!,n)\simeq s\,C(2^-;3) {-1 \over |{q_{4}^\perp}|^2}\,  V(q_{4}; 4,\ldots, n{-}1; q_{n})\, {-1 \over |{q_{n}^\perp}|^2}\, C(1^-; n)\,,
\end{align}
where the generalised Lipatov vertices admit the following CHY-type representation,
\begin{align}\label{Lipatov-main}
\begin{aligned}
  V\big(q_4; 4,\ldots, n{-}1; & q_n\big)
  \\
  \,=\, ({q_4^\perp})^\ast {q_n^\perp} \, \bigintsss &\prod\limits_{a=4}^{n-1} {d\sigma_a dt_a \over t_a}\,
  {1 \over \sigma_{45}\cdots\sigma_{n-2, n-1}\sigma_{n-1}}\,
  \left( \prod_{i\in\mathfrak{P}, I\in{\mathfrak N}} {k_I^\perp \over k_i^\perp}  \right)
  \\
  \times&\prod_{I\in{\mathfrak N}} 
  \delta\Bigg( {k_I^\perp}  - \sum_{i\in\mathfrak{P}} {t_i t_I \over \sigma_I {-} \sigma_i} {k_i^+} 
  + {t_I \over 1 - \sum_{J\in{\mathfrak N}} t_J}\, q_4^\perp \Bigg)
  \\
  \times&\prod_{I\in{\mathfrak N}} 
  \delta\Bigg( (k_I^\perp)^\ast 
  - {k_I^+ \over k_I^\perp}\sum_{i\in\mathfrak{P}} {t_i t_I \over \sigma_I {-} \sigma_i} ({k_i^\perp})^\ast 
  - {\zeta_I  \over  1 + \sum_{J\in{\mathfrak N}} \zeta_J}\, ({q_n^\perp})^\ast \Bigg)
  \\
  \times&\prod_{i\in\mathfrak{P}} 
  \delta\Bigg( 1  - \sum_{I\in{\mathfrak N}} {t_i t_I \over \sigma_i {-} \sigma_I} {k_I^+  \over  k_I^\perp}  + t_i\Bigg)
  \delta\Bigg( 1 - {k_i^+ \over k_i^\perp} \sum_{I\in{\mathfrak N}} {t_i t_I \over \sigma_i {-} \sigma_I}  + \zeta_i \Bigg)\,.
\end{aligned}
\end{align}
and $\mathfrak{N}$ and $\mathfrak{P}$ denote the subsets of $\{4,\ldots,n{-}1\}$ of particles with negative and positive helicity respectively.
We have checked that this CHY-type formula is consistent with known results in the literature for generalised Lipatov vertices for the production up to three particles, as well as for NMHV-type Lipatov vertices for the production of up to four particles. In addition, we show in Appendix~\ref{app:factorisations} that eq.\,\eqref{Lipatov-main} has the correct factorisation properties. In particular, if $y_4$ ($y_{n-1}$) is much greater (smaller) than the rapidities of all the other emitted particles, then the Liptatov vertex itself factorises,
\begin{align}\label{eq:V_r_V}
  V\big(q_4&; 4,\ldots, n{-}1;  q_n\big) 
  \\
  &\simeq\,  \left\{\begin{array}{ll}
  \displaystyle
   V\big(q_4; 4;  q_5\big)\,{-1\over |q_5^\perp|^2}\,  V\big(q_5; 5,\ldots, n{-}1;  q_n\big)\,, &  y_4\gg y_i\,,\,\, 5\le i \le n{-}1,\\
   \displaystyle
   V\big(q_4; 4,\ldots, n{-}2;  q_{n-1}\big)\,{-1\over |q_{n-1}^\perp|^2}\,   V\big(q_{n-1}; n{-}1;  q_n\big)\,, &  y_{n-1}\ll y_i,\,\, 4\le i \le n{-}2.
 \end{array}\right.
 \nonumber
\end{align}
Using this property we can iterate the factorisation and gradually approach the quasi-multi-Regge limit $y_3\gg\cdots\gg y_r  \simeq\cdots\simeq y_s\gg  \cdots \gg y_n$, and we see that the amplitude takes the factorised form in eq.\,\eqref{eq:Amp_V_QMRK}. We emphasise again the factorisation is a direct consequence of Conjecture 1 and the CHY representation for gluon amplitudes. In particular, we find that this factorisation holds for arbitrary helicity configurations and that the ensuing generalised Lipatov vertices are universal and do not depend on the quantum numbers of the other particles involved in the scattering.

\subsection{Other types of quasi-multi-Regge limits}
In the previous section we have only studied some very specific quasi-multi-Regge limits, and we have shown that in those cases Conjecture 1 implies that the amplitude has the expected factorisation into universal building blocks. In principle, one could also consider more general quasi-multi-Regge limits, like for example
\beq\label{eq:general_QMRL_example}
y_3\simeq\ldots\simeq y_{r-1}\gg y_{r}\simeq \ldots\simeq y_{s}\gg y_{s+1}\simeq\ldots\simeq y_n\,.
\eeq
In this limit the amplitude is expected to factorise as 
\begin{align}\label{eq:general_QMRK_example}
  {\cal A}_n(1,\!\ldots\!,n)\simeq s\,C(2;3,\ldots,r-1) {-1 \over |{q_{r}^\perp}|^2}  V(q_{r}; r,\ldots,s; q_{s+1}) {-1 \over |{q_{s+1}^\perp}|^2} C(1;s{+}1,\ldots, n)\,.
\end{align}

Unfortunately, we are currently not able to derive this factorisation from Conjecture 1 and the CHY representation of the amplitude. The main obstacle is that for these more general quasi-multi-Regge limits we cannot identify a set of variables that enter the scattering equations linearly in the limit. This property, however, was the cornerstone in previous sections to prove factorisation in (quasi-)multi-Regge limits. We stress that our inability to derive factorisations for more general QMRKs from Conjecture 1 and the CHY representation of the amplitude does by no means imply that no such factorisation exists. Indeed, if we consider for example eq.\,\eqref{eq:general_QMRK_example} and we insert the CHY-type representation for the generalised impact factors and Lipatov vertices in eqs.~\eqref{Impact-factor-main} and~\eqref{Lipatov-main} we obtain a representation for the amplitude in this limit that is consistent with the known factorisations of the amplitude in all soft, collinear and multi-Regge limits. We then find it hard to imagine that the amplitude could take any other form in this limit than the one given in eq.\,\eqref{eq:general_QMRK_example}.

Finally, let us conclude this section by commenting on amplitudes involving quarks. There are various quasi-multi-Regge limits which involve one or more quark pairs in the final state. 
For example, in the quasi-multi-Regge limit $y_3\gg y_4\simeq y_5\gg y_6$, we have
\begin{align}
  {\cal A}_6(1,2,3,4_q,5_{\bar{q}},6) \,\simeq\, s\,C(2;3) {-1 \over |{q_{4}^\perp}|^2}  V(q_{4}; 4_q,5_{\bar{q}}; q_{6}) {-1 \over |{q_{6}^\perp}|^2} C(1;6)\,.
\end{align}
Since there are CHY representations for tree-level amplitudes involving massless quarks~\cite{He:2016dol}, and since the scattering equations are universal and do not depend on the details of the theory, we can immediately extend our analysis to these amplitudes and use Conjecture 1 to prove their factorisation in various quasi-multi-Regge limits. In particular, we can obtain CHY-type representations for the corresponding generalised impact factors and Lipatov vertices, and we have checked that also in this case we are able to reproduce the known analytic expressions from the literature~\cite{DelDuca:1995zy,DelDuca:1996nom,DelDuca:1999iql,Duhr:2009uxa}.

\section{Conclusion}
\label{sec:conclusion}
In this paper we have initiated the study of Regge kinematics through the lens of the scattering equations and the CHY formula. Based on numerical studies, we have formulated a precise conjecture about the behaviour of the solutions to the scattering equations in a Regge limit, both in their $D$ and four-dimensional guises. While we currently have no proof of our conjecture, we have tested its validity by showing that we can derive the expected factorisation of the amplitudes when we combine the conjecture with the four-dimensional CHY formula. This is a highly non-trivial prediction of our conjecture which gives us confidence that it describes the correct asymptotic behaviour of the solutions in QMRK. 

Our conjecture is not only of formal interest, but it has concrete applications to tree-level scattering amplitudes. In particular, we have applied our conjecture to show that in MRK the four-dimensional scattering equations have a unique solution (up to $GL(2,\mathbb{C})$ redundancy), independently of the multiplicity, and we have explicitly determined this MRK solution. We find it remarkable that in MRK it is possible to find an exact solution of the four-dimensional scattering equations for arbitrary multiplicities. Indeed, so far this has only been achieved for the MHV sector and in a special eikonal Regge kinematical regime. In QMRK we cannot obtained exact solutions to the scattering equations anymore. Instead we have derived CHY-type formulas for the generalised impact factors and Lipatov vertices valid for an arbitrary number of particles, and we have checked that these formulas reproduce known analytic results from the literature for low multiplicities.

We see two possible directions for future research. First, it would be interesting to find a proof of our conjecture. This would not only clarify some new mathematical property of the scattering equations, but it would have direct implications for tree-level scattering amplitudes. Indeed, so far the factorisation of colour-ordered helicity amplitudes has only been rigorously proven for arbitrary multiplicities for the simplest helicity configurations. In this paper we have shown that our conjecture implies the expected factorisation for arbitrary helicity configurations. A rigorous mathematical proof of our conjecture would thus immediately lead to an elegant proof of the Regge factorisation of all tree-level amplitudes in gauge theories. 


Second, while in this paper we have focused exclusively on gluon amplitudes, the scattering equations and the CHY formula are valid for much larger classes of massless quantum field theories, and thus our conjecture applies also to theories other than Yang-Mills. In particular, since we have shown that our conjecture implies the factorisation of gauge theory amplitudes in various quasi-multi-Regge limits, it would be interesting to investigate if similar factorisations can be derived for other massless quantum field theories in QMRK. The study of Regge kinematics has so far mostly focused on gauge theory amplitudes and gravity~\cite{Lipatov:1982vv,zhengwen_gravity}, and it would be interesting to study the implications for other theories. We leave this study for future work.

\section*{Acknowledgements}
The authors are grateful to Vittorio Del Duca and Song He for useful discussions and a careful reading of the manuscript. ZL thanks the CERN TH Department for hospitality during various stages of this work. CD and ZL acknowledge the support and the hospitality of the Galileo Galilei Institute in Florence during the final phase of this project. This work is supported by the ERC grant 637019 ``MathAm'' and the ``Fonds Sp\'ecial de Recherche'' (FSR) of the UCLouvain.

\appendix
\section{From $D$ to 4 dimensions}
\label{app:4dSE}
In this appendix, we clarify the relation between $D$- and four-dimensional scattering equations. 
Let us start with a rational map from the moduli space ${\mathfrak M}_{0,n}$ to momentum space as follows \cite{Cachazo:2013iaa, Cachazo:2013gna}:
\begin{align}\label{SE-map}
  k_a^\mu \,&=\, {1 \over 2\pi i}\oint\limits_{|z-\sigma_a|=\varepsilon}\, dz\, w^\mu(z),
\end{align}
with
\begin{align}\label{}
  w^\mu(z) \,=\, \sum_{b=1}^n {k_b^\mu \over z-\sigma_b} \,=\, {P^\mu(z) \over \prod_{b=1}^n (z-\sigma_b)},
\end{align}
where $\sigma_a$ denote the marked points in ${\mathfrak M}_{0,n}$.
Clearly, $P^\mu(z)$ is a polynomial of degree $n{-}2$.
Moreover, it turned out that $P^\mu(z)$ requires to be {\it null}.
The meromorphic function
\begin{align}\label{}
  w(z)^2 \,\equiv\, w_\mu(z) w^\mu(z) \,=\,\sum_{a} {1 \over z-\sigma_a} \sum_{b\ne a} {2k_a\cdot k_b \over \sigma_a - \sigma_b},
\end{align}
has no pole at $z=\infty$ and only simple pole at each $z=\sigma_a$, thus using the residue theorem one can immediately get the constraints to determine the marked points $\sigma_a$, i.e.,
\begin{align}\label{ap-SE-chy-1}
  0 \,=\, {1 \over 4\pi i}\oint\limits_{|z-\sigma_a|=\varepsilon}\, dz\, w(z)^2 \,=\, \sum_{b\neq a} {k_a\cdot k_b \over \sigma_{a} - \sigma_{b}},
  \qquad a=1, 2, \ldots, n,
\end{align}
which are just the {scattering equations}.
This system of algebraic equations owns a global ${SL}(2,{\mathbb C})$ symmetry, and thus only $n{-}3$ out of the $n$ equations are independent.

In four dimensions the null vector $P^\mu(z)$ can be rewritten in spinor variables as follows:
\begin{align}\label{SE-map-P}
  P^{\alpha\dot\alpha}(z) \,&=\, 
  \left(\prod_{a=1}^n (z-\sigma_a)\right) \sum_{b=1}^n {\lambda_b^\alpha\tilde\lambda_b^{\dot\alpha} \over z-\sigma_b}
  \,=\,\lambda^\alpha(z)\tilde\lambda^{\dot\alpha}(z),
\end{align}
where $\lambda(z)$ and $\tilde\lambda(z)$ are polynomials in $z$ of degree $d$ and $\tilde d$ respectively, $d, \tilde d \in\{1,\ldots, n{-}3\}$ and $d+\tilde{d} = n{-}2$.
The polynomials $\lambda(z)$ and $\tilde\lambda(z)$ can be constructed in the following way:
\begin{align}\label{SE-map-spinors}
  \lambda^\alpha(z) \,&=\, \prod_{a\in \mathfrak{N}} (z-\sigma_a)\, \sum_{I\in \mathfrak{N}} \frac{ t_I \lambda^\alpha_I}{z-\sigma_I}, \qquad
  \lambda^{\dot\alpha}(z) \,=\, \prod_{a\in\mathfrak{P}} (z-\sigma_a)\, \sum_{i\in\mathfrak{P}} {t_i\tilde\lambda_i^{\dot\alpha}\over z-\sigma_i}.
\end{align}
Here $\mathfrak{N}$ is any subset of $\{1,\ldots,n\}$ with length $2\le k \le n{-}2$ and $\mathfrak{P}$ is the corresponding complement.
Therefore, $\lambda(z)$ has degree $d=k{-}1$, while $\tilde\lambda(z)$ has degree $\tilde{d}=n {-} k {-} 1$.
Here the variables $t_a$ with $a\in\{1,2,\ldots, n\}$ can be understood as the combinations of coefficients in polynomials $\lambda(z)$ and $\tilde\lambda(z)$ in eq.\,\eqref{SE-map-P}. Plugging \eqref{SE-map-spinors} into eq.~\eqref{SE-map-P}, we find
\begin{align}
  P^{\alpha\dot\alpha}(z) \,&=\, \lambda^\alpha(z)\tilde\lambda^{\dot\alpha}(z)  
  \nonumber\\
  \,&=\, \prod_{a=1}^n (z-\sigma_a)\, \left(\sum_{I\in\mathfrak{N}} \frac{ t_I \lambda^\alpha_I}{z-\sigma_I} \right)
  \left( \sum_{i\in \mathfrak{P}} {t_i\tilde\lambda_i^{\dot\alpha}\over z-\sigma_i} \right)
  \nonumber\\
  \,&=\, \prod_{a=1}^n (z-\sigma_a)\, 
  \sum_{I\in\mathfrak{N}}  \sum_{i\in\mathfrak{P}} t_I t_i \lambda^\alpha_I\tilde\lambda_i^{\dot\alpha}
  \left( {1 \over z-\sigma_I} - {1 \over z-\sigma_i} \right)\, {1 \over \sigma_I-\sigma_i}
  \nonumber\\
  \,&=\, \prod_{a=1}^n (z-\sigma_a)\, \left(
  \sum_{I\in\mathfrak{N}}  {\lambda^\alpha_I \over z-\sigma_I} \sum_{i\in\mathfrak{P}} 
  {\tilde\lambda_i^{\dot\alpha} \over (I\,i)}
  + \sum_{i\in\mathfrak{P}} {\tilde\lambda_i^{\dot\alpha} \over z-\sigma_i}
  \sum_{I\in\mathfrak{N}} {\lambda^\alpha_I \over (i\,I)}
  \right)
\end{align}
which implies that
\begin{align}\label{Appendix-SE-4d-twistor-app}
  \tilde\lambda_I^{\dot\alpha} \,=\, \sum_{i\in\mathfrak{P}} {\tilde\lambda_i^{\dot\alpha} \over (I\,i)},
  \quad I\in \mathfrak{N};
  \qquad
  \lambda_i^{\alpha} \,=\, \sum_{I\in\mathfrak{N}} {\lambda^\alpha_I \over (i\,I)},
  \quad i\in\mathfrak{P}\,.
\end{align}
We see that that in four dimensions the scattering equations fall into different sectors characterized by $k\in\{2,\ldots, n{-}2\}$.
The number of solutions of the scattering equations in sector $k$ is given by the Eulerian number $\eulerian{n{-}3}{k{-}2}$~\cite{Cachazo:2013iaa, Spradlin:2009qr}. It is well-known that $\sum_{k=2}^{n-2} \eulerian{n-3}{k-2} = (n{-}3)!$~\cite{Petersen:2015Eu}, in agreement with the fact that the scattering equations in eq.~\eqref{SE-chy-1} have $(n{-}3)!$ independent solutions.
There is an overall rescaling redundancy, i.e.~$\big\{\lambda(z), \tilde\lambda(z)\big\} \to \big\{c \lambda(z), c^{-1}\tilde\lambda(z)\big\}$ leaves $P(z)$ invariant, and so this system of equations has a global ${GL(2,{\mathbb C})}= {SL(2,{\mathbb C})}\times {GL(1,{\mathbb C})}$ symmetry. As a byproduct we have shown that the solutions $\sigma_a$ solve at the same time the $D$- and four-dimensional scattering equations.


\section{The asymptotic behaviour of the four-dimensional scattering equations}
\label{app:conjecture}

\subsection{The conjecture for all helicity sectors}
In this appendix, we present the conjecture on the behavior of the solutions of the four-dimensional scattering equations in QMRK for all helicity sectors. We conjecture that if we fix the $SL(2,\mathbb{C})$ redundancy according to eq.\,\eqref{gauge-fixing-d-se}, i.e.,
\begin{align*}\label{}
  \sigma_1 \,=\, 0, \quad \sigma_2 \,=\, t_2 \,\to\, \infty, \quad t_1 \,=\, -1\,,
\end{align*}
 then the solutions of four-dimensional scattering equations behave as
\begin{align}\label{4d-conjecture-general}
\begin{aligned}
  \sigma_a \,&=\, {\cal O}\Big(k_a^+\,\kappa^{(h_n - h_3)/2}\Big), \qquad 3\le a\le n,
  \\
  t_i \,&=\, {\cal O}\Big(\sqrt{k_i^+ \,\kappa^{(1-h_3-h_i)}}\Big), \quad i\in{\mathfrak P}, i\ne 3,
  \\
  t_I \,&=\, {\cal O}\Big(\sqrt{k_I^+ \,\kappa^{(-h_I+h_n-1)}}\Big), \quad I\in\overline{\mathfrak N}, I\ne n,
  \\
  t_3 \,&=\, {\cal O}\Big(\sqrt{k_3^+ \,\kappa^{(1-h_3-h_3)}}\Big),
  \qquad
  t_n \,=\, {\cal O}\Big(\sqrt{k_n^+ \,\kappa^{(1-h_3-h_n)}}\Big).
\end{aligned}
\end{align}
Obviously, in the case $\{3,n\} \subseteq \mathfrak{P}$, $h_3=h_n=1$, it immediately goes back to Conjecture 1
\begin{align}\label{}
  \sigma_a \,=\, {\cal O}\left( {k_a^+} \right),\qquad 
  t_a \,=\, {\cal O}\left(\!\!\sqrt{k_a^+\,\kappa^{-h_a}}\right)\,,
\end{align}

Here we focus on the case of $3\in\overline{\mathfrak N}$ and/or $n\in\overline{\mathfrak N}$.
We have performed a detailed numerical analysis for these cases up to seven points.
To be clear, let us see a simple example of five points.
We consider $\mathfrak{P}=\{3,4\}$, and in this case it is easy to obtain the unique exact solution as follows:
\begin{align}\label{}
  \sigma _3 = {[{24}] \over [{14}]},~
  t_3 = {[{24}] \over [{34}]},~
  \sigma _4 = {[{23}] \over [{13}]},~
  t_4 = -{[{23]} \over [{34}]},~
  \sigma _5 = {[{15}] [{23}] [{24}] \over [{13}] [{14}] [{25}]},~
  t_5 = {[{12}] [{35}] [{45}] \over [{13}] [{14}] [{25}]}.
\end{align}
In the MRK, we have
\begin{align}\label{}
\begin{aligned}\label{}
  \sigma_3 \,&=\, - {(k_4^\perp)^\ast \over {k_4^+}} \big(k_3^+ \kappa^{-1}\big), 
   \\
   t_3 \,&=\, -1, 
\end{aligned}
\quad
\begin{aligned}\label{}
   \sigma_4 \,&=\,- {(k_3^\perp)^\ast \over {k_4^+}} \big({k_4^+} \kappa^{-1}\big), 
   \\
   t_4 \,&=\, {(k_3^\perp)^\ast  \over (k_4^\perp)^\ast}\, \sqrt{k_4^+ \kappa^{-1}}, 
\end{aligned}
\quad
\begin{aligned}\label{}
   \sigma_5 \,&=\,- {(k_3^\perp k_4^\perp)^\ast \over {k_4^+} (k_5^\perp)^\ast} \big(k_4^+ \kappa^{-1}\big),
   \\
   t_5 \,&=\, \sqrt{(k_5^\perp)^\ast \over k_5^\perp}.
\end{aligned}
\end{align}
This exactly matches our conjecture presented above.

\subsection{Helicity conservation}
Here we show that our conjecture implies that helicity is conserved by the impact factors $C(1;n)$ and $C(2;3)$.
We study the case of $n\in\mathfrak{N}$ in the limit $y_n\ll y_i$ ($3\le i < n$).
In this case, all equations in eq.\,\eqref{4dSE-lightcone} are independant of $\sigma_n$ at leading order, since in the limit $y_n\ll y_i$ we have
\begin{align}\label{}
   {1 \over \sigma_i - \sigma_n} \,\simeq\, {1 \over \sigma_i}.
\end{align}
Therefore, the integral over $\sigma_n$ vanishes at leading order, which implies $C(1^-;n^-)=0$.
This is consistent with the fact the helicity is conversed by leading impact factors.
A similar analysis can be performed to show $C(2^-;3^-)=0$. 

Alternatively, we can show that the impact factor $C(1^-;n^-)$ vanishes by using the scattering equation with $n\in\mathfrak{P}$.
One can show that the amplitudes ${\cal A}_n(1^-,2^-,\ldots,i^+,\ldots,n^-)$ and ${\cal A}_n(1^-,2^-,\ldots,i^-,\ldots,n^+)$ can be computed using the same CHY formula in eq.~\eqref{amp-YM-4d}, up to adding a additional factor to the integrand, i.e.,
\beq\label{}
  {1 \over (12)\ldots(n1)} \,\longrightarrow\, {1 \over (i\,n)^4}\,\times {1 \over (12)\ldots(n1)}\,.
\eeq
In the limit $y_n\ll y_i$ with $3\le i < n$, we have
\beq\label{}
  {1 \over (i\,n)^4} \,\simeq\, \left({t_i t_n \over \sigma_i}\right)^4 
  \,=\, {\cal O}\bigg(\left({k_i^+ \over k_n^+}\right)^2\bigg).
\eeq
This shows that the amplitude ${\cal A}_n(1^-,2^-,\ldots,n^-)$ is suppressed in the limit $y_i\gg y_n$.


\section{An alternative formula for Lipatov vertices}
\label{app:Lipatov2}

In previous analyses, we always remove the four equations $\bar{\cal S}^{\dot\alpha}_1 = \bar{\cal S}^{\dot\alpha}_2 = 0$ and identify them with the momentum conservation constraint, cf.~eq.\,\eqref{SEs-12-momentum-conserv}.
In the limit $y_3\gg y_4 \simeq \cdots\simeq y_{n-1}\gg y_n$, we find that it is convenient to use the equations $\bar{\cal S}^{\dot\alpha}_1 = \bar{\cal S}^{\dot\alpha}_2 = 0$ to localize the integrals over $\{\sigma_3,\tau_3,\sigma_n,\tau_n\}$.
We can then identify $\delta^2({\cal S}^{\alpha}_3) \delta^2({\cal S}^{\alpha}_n)$ with the $\delta$-function expressing momentum conservation:
\begin{align}
  \delta^2\big( {\cal S}^{\alpha}_3 \big)\, \delta^2\big( {\cal S}^{\alpha}_n \big) 
  \,=\, s\, k_3^\perp (k_n^{\perp})^\ast\, \delta^4\left(\sum_{a=1}^{n} k_a^\mu \right),
\end{align}
Finally, this leads to the following alternatively equivalent formula for the generalized Lipatov vertex: 
\begin{align}\label{Lipatov-main-2}
\begin{aligned}
  V\big(q_4; 4,\ldots, n{-}1; & q_n\big)
  \\
  \,=\,  ({q_4^\perp}^\ast {q_n^\perp}) \bigintsss & 
  \prod\limits_{a=4}^{n-1} {d\sigma_a  d\tau_a \over \tau_a}\, {{\cal J}_{1,2;3,n} \over \sigma_{45}\cdots\sigma_{n{-}2, n{-}1} \sigma_{n{-}1}}\,
  \Bigg( \prod_{i\in\mathfrak{P}, I\in\mathfrak{N}}  {k_I^\perp  \over  k_i^\perp} \Bigg)
  \\
  \times &
  \prod_{I\in\mathfrak{N}}
  \delta\Bigg(
  {k_I^\perp}  - \sum_{i\in\mathfrak{P}} \left({\tau_I \tau_i \over \sigma_I {-} \sigma_i} + {\tau_I \tau_i \over \sigma_i} \right)  {k_i^+}  - \tau_I {q_n^\perp}
  \Bigg)
  \\
  \times &
  \prod_{I\in\mathfrak{N}}
  \delta\Bigg(
  (k_I^\perp)^\ast  - \sum_{i\in\mathfrak{P}} \left( {k_I^+ \over k_I^\perp} {\tau_I \tau_i  \over \sigma_I {-} \sigma_i} - \zeta_I \tau_i \right) ({k_i^\perp})^\ast
  - \zeta_I ({q_4^\perp})^\ast
  \Bigg)
  \\
  \times &
  \prod_{i\in\mathfrak{P}}
  \delta\Bigg(1  -  \sum_{I\in\mathfrak{N}} {\tau_i \tau_I \over \sigma_i {-} \sigma_I} {k_I^+  \over  k_I^\perp}   +  \tau_i \Bigg)
  \delta\Bigg( 1   -  {k_i^+ \over k_i^\perp} \sum_{I\in\mathfrak{N}} {\tau_i \tau_I \over \sigma_i {-} \sigma_I}   +  \zeta_i \Bigg),
\end{aligned}
\end{align}
where
\begin{align}\label{Jacobian-123n}
  {\cal J}_{1,2; 3,n} \,&=\,
  {q_4^\perp} ({q_n^\perp})^\ast\,
  \bigg( {q_n^\perp}  +  \sum_{i\in\mathfrak{P}} \zeta_i {k_i^\perp}  \bigg)^{-1}
  \bigg( ({q_4^\perp})^\ast  - \sum_{i\in\mathfrak{P}} \tau_i ({k_i^\perp})^\ast \bigg)^{-1}.
\end{align}

\section{Factorisations of impact factors and Lipatov vertices}
\label{app:factorisations}
In this appendix we show that the generalised impact factors and Lipatov vertices that appear in QMRK have the expected factorisations in soft, collinear and Regge limits.

\subsection{Soft limits}

Let us start by discussing the soft limits of the impact factors and the Lipatov vertices defined through the CHY-type formulas in eqs.~\eqref{Impact-factor-main} and~\eqref{Lipatov-main}. The argument is the same in both cases and follows closely the analysis of the soft limit of the full amplitude in ref.~\cite{Cachazo:2013hca,Cachazo:2013iea}. In the following we only discuss the case of the generalised Lipatov vertex $V(q_4;4,\ldots,n{-}1;q_n)$.

Without loss of generality we assume that $k_5\to0$, and we assume that the soft gluon has positive helicity. The argument in other cases is identical.
Keeping only the leading behaviour as $k_5\to0$ in eq.~\eqref{Lipatov-main}, we can write
\begin{align}\label{Lipatov-soft-1}
  V\big(q_4; 4,\ldots, n{-}1; q_n\big)
  \,\simeq\,  {1 \over k_5^\perp} \bigintsss\prod\limits_{a=4}^{n-1} {d\sigma_a d\tau_a \over \tau_a}\,\left({1 \over \sigma_{45}} + {1 \over \sigma_{56}}\right)
   \delta^2\big({S}^\alpha_5\big)\,
    {\cal I}_{n-5}\,,
\end{align}
where ${\cal I}_{n-5}$ collects the terms in the integrand of eq.~\eqref{Lipatov-main} independent of $\sigma_5$ and $\tau_5$,
\begin{align}\label{}
\begin{aligned}
  {\cal I}_{n-5} \,=~ &
  {({q_4^\perp})^\ast {q_n^\perp} \over \sigma_{46}\sigma_{67}\cdots\sigma_{n-2, n-1}\sigma_{n-1}}\,
  \Bigg( \prod_{i\in\mathfrak{P}\backslash\{5\}, I\in{\mathfrak N}} {k_I^\perp \over k_i^\perp}  \Bigg)
  \\
  &\times\prod_{I\in{\mathfrak N}} 
  \delta\Bigg( {k_I^\perp}  - \sum_{i\in\mathfrak{P}\backslash\{5\}} {\tau_i \tau_I \over \sigma_I {-} \sigma_i} {k_i^+} 
  + {\tau_I \over 1 - \sum_{J\in{\mathfrak N}} t_J}\, q_4^\perp \Bigg)
  \\
  &\times\prod_{I\in{\mathfrak N}} 
  \delta\Bigg( (k_I^\perp)^\ast  - {k_I^+ \over k_I^\perp}\sum_{i\in\mathfrak{P}\backslash\{5\}} {\tau_i \tau_I \over \sigma_I {-} \sigma_i} ({k_i^\perp})^\ast 
  -  {\zeta_I  \over  1 + \sum_{J\in{\mathfrak N}} \zeta_J}\, ({q_n^\perp})^\ast \Bigg)
  \\
  &\times\prod_{i\in\mathfrak{P}\backslash\{5\}} 
  \delta\Bigg(  1  - \sum_{I\in{\mathfrak N}} {\tau_i \tau_I \over \sigma_i {-} \sigma_I} {k_I^+  \over  k_I^\perp}  + \tau_i\Bigg)
  \delta\Bigg( 1 -  {k_i^+ \over k_i^\perp}\sum_{I\in{\mathfrak N}} {\tau_i \tau_I \over \sigma_i {-} \sigma_I}  + \zeta_i \Bigg).
\end{aligned}
\end{align}
Let us write down ${S}^\alpha_5$ explicitly:
\beq\bsp
  \label{soft-SE-51}
  {S}^1_5 \,&=\,  1 + \tau_5 \left(1 - \sum_{I\in{\mathfrak N}} {\tau_I \over \sigma_5 {-} \sigma_I} {k_I^+  \over  k_I^\perp}\right),
  \\
  {S}^2_5 \,&=\,  1 +  \tau_5 {k_5^+ \over k_5^\perp} \left({1 \over \sigma_5} - \sum_{I\in{\mathfrak N}} {\tau_I \over \sigma_5 {-} \sigma_I}\right).
\esp\eeq
As a first step, let us use the delta function $\delta({S}^1_5)$ in eq.\,\eqref{soft-SE-51} to fix $\tau_5$. We then obtain
\begin{align}\label{Lipatov-soft-2}
  V\big(q_4; 4,\ldots, n{-}1; q_n\big)
  \,\simeq\,  {1 \over k_5^\perp} \bigintsss\prod\limits_{a=4,a\ne 5}^{n-1} {d\sigma_a d\tau_a \over \tau_a}\,{\cal I}_{n-5}
   \oint_{\cal C} {d\sigma_5 \over {S}^2_5}\,\left({1 \over \sigma_{45}} + {1 \over \sigma_{56}}\right),
\end{align}
where the contour ${\cal C}$ is defined to encircle the zeros of ${S}^2_5$, which is given by
\begin{align}
  \label{soft-SE-52-2}
  {S}^2_5\big|_{{S}^1_5=0} \,&=\,  1 \,-\, {k_5^+ \over k_5^\perp}\left({1 \over \sigma_5} - \sum_{I\in{\mathfrak N}} {\tau_I \over \sigma_5 {-} \sigma_I}\right)
  \left(1 - \sum_{I\in{\mathfrak N}} {\tau_I \over \sigma_5 {-} \sigma_I} {k_I^+  \over  k_I^\perp}\right)^{-1}.
\end{align}
We apply the global residue theorem, which allows us to express the residue at ${S}^2_5=0$ in terms of the residues at $\sigma_5=\sigma_4$ and $\sigma_5=\sigma_6$.
Even though the computation depends on the helicity configuration of the two adjacent legs, the final result does not, and a straightforward calculation gives
\begin{align}\label{Lipatov-soft-factor}
      V\big(q_4; 4,5,\ldots, n{-}1; q_n\big)
      \,\simeq\,   
      {\braket{4\,6} \over \braket{4\,5} \braket{5\,6}}\,V\big(q_4; 4,6,\ldots, n{-}1; q_n\big).
\end{align}


\subsection{Collinear limit}
We now study the generalised impact factors and Lipatov vertices in the limit where a pair of produced particles become collinear. We again restrict ourselves to the study of Lipatov vertices, and we assume that the momenta $k_4$ and $k_5$ are collinear. 
We take the following parametrization for the two collinear massless momenta \cite{Stieberger:2015kia},
\begin{align}\label{collinear}
\begin{aligned}
  {k_4^+} \,&=\,  z {k_x^+} - 2\epsilon \sqrt{z(1 {-} z)} \big({k_x^+}{k_y^+}\big)^{1/2}  + \epsilon^2 (1 {-} z) {k_y^+},
  \\
  {k_4^-} \,&=\,  z {k_x^-}
  - \epsilon \sqrt{z(1 {-} z)} \big(({k_x^\perp})^\ast  {k_y^\perp}+{k_x^\perp} ({k_y^\perp})^\ast\big) \big({k_x^+}{k_y^+}\big)^{-1/2}
  + \epsilon^2 (1 {-} z) {k_y^-},
  \\
  {k_4^\perp} \,&=\,  z {k_x^\perp}
  - \epsilon \sqrt{z(1 {-} z)} \big({k_x^\perp}{k_y^+} + {k_x^+} {k_y^\perp}\big) \big({k_x^+}{k_y^+}\big)^{-1/2}
  + \epsilon^2 (1 {-} z) {k_y^\perp},
  \\
  {k_5^+} \,&=\,  (1 {-} z) {k_x^+}
  + 2\epsilon \sqrt{z(1 {-} z)} \big({k_x^+}{k_y^+}\big)^{1/2}
  + \epsilon^2 z {k_y^+},
  \\
  {k_5^-} \,&=\,  (1 {-} z) {k_x^-}
  + \epsilon \sqrt{z(1 {-} z)} \big(({k_x^\perp})^\ast  {k_y^\perp}+{k_x^\perp} ({k_y^\perp})^\ast\big) \big({k_x^+}{k_y^+}\big)^{-1/2}
  + \epsilon^2 z {k_y^-},
  \\
  {k_5^\perp} \,&=\,  (1 {-} z) {k_x^\perp}
  + \epsilon \sqrt{z(1 {-} z)} \big({k_x^\perp}{k_y^+} + {k_x^+} {k_y^\perp}\big) \big({k_x^+}{k_y^+}\big)^{-1/2}
  + \epsilon^2 z {k_y^\perp}.
\end{aligned}
\end{align}
The collinear limit is realised as $\epsilon\to 0$ and the collinear direction is $K=k_4 + k_5 = k_x + {\cal O}(\epsilon^2)$.

In ref.~\cite{Nandan:2016ohb} it was shown that in the limit where two gluons become collinear only those solutions contribute where $\sigma_4$ and $\sigma_5$ coincide in the limit. It is then useful to perform the change of variables from $(\sigma_4,\sigma_5)$ to $(\rho,\xi)$:
\begin{align}\label{collinear-variables}
  \sigma_4 \,=\, \rho + {1 \over 2}\epsilon\xi + {\cal O}(\epsilon^2)\,, \qquad
  \sigma_5 \,=\, \rho -  {1\over 2}\epsilon\xi + {\cal O}(\epsilon^2)\,,
\end{align}
such that that $\sigma_4 {-} \sigma_5 = \epsilon\xi + {\cal O}(\epsilon^2)$ and ${d\sigma_4}{d\sigma_5}=\epsilon{d\xi} {d\rho}$.
%
%
%
%
In these new variables eq.~\eqref{Lipatov-main} takes the form:
\begin{align}\label{}
  V\big(q_4; 4,5, & \ldots, n{-}1; q_n\big)
  \,=\, ({q_4^\perp})^\ast {q_n^\perp}\,
  \Bigg( \prod_{i\in\mathfrak{P}, I\in{\mathfrak N}} {k_I^\perp \over k_i^\perp}  \Bigg)
  \nonumber\\
  &\times
  \bigintsss \prod\limits_{a=6}^{n-1} {d\sigma_a d\tau_a \over \tau_a}\,
  \bigintsss {d\xi d\rho d\tau_4 d\tau_5 \over \xi \tau_4 \tau_5}\,
  {\prod\limits_{I\in{\mathfrak N}} \delta^2\big( {S}^{\dot\alpha}_I \big) \prod\limits_{i\in\mathfrak{P}} \delta^2\big( {S}^\alpha_i \big) 
  \over (\rho {-} \sigma_6) \sigma_{67} \cdots\sigma_{n-2, n-1}\sigma_{n-1}}.
  \label{Lipatov-collinear-main}
\end{align}
We now show that after integrating out $\xi$ and some linear combination of $\tau_4$ and $\tau_5$ we can recover the expected factorised form of the Lipatov vertex in the collinear limit. In order to proceed, we need to analyse separately the cases where the collinear particles have either the same or opposite helicities.

\subsubsection*{$(h_4,h_5)=(+,+)$}
Let us first consider the case where the two particles have positive helicity.
It is convenient to perform the change of variables from $(\tau_4,\tau_5)$ to $(\tau_x,\tau_y)$ defined by,
\begin{align}
\begin{aligned}
  \tau_x \,&=\, z\,\tau_4 + (1{-}z)\,\tau_5     \textrm{~~and~~}\tau_y \,&=\, \sqrt{z(1 {-} z)}\, \big(\tau_5-\tau_4\big) \,.
\end{aligned}
\end{align}
such that
\begin{align}
  d\tau_4 d\tau_5 \,=\, {1 \over \sqrt{z (1 {-} z)}} d\tau_x d\tau_y\,.
\end{align}
Equation~\eqref{Lipatov-collinear-main} then becomes,
\beq\bsp
  V\big(q_4&; 4,5, \ldots, n{-}1; q_n\big)
  \,=\,  {(q_4^\perp)^\ast {q_n^\perp} \over (k_x^\perp)^2}\, {1 \over (z(1 {-} z))^{3/2}}
  \Bigg( \prod_{i\in\mathfrak{P}\backslash\{4,5\}, I\in{\mathfrak N}} {k_I^\perp \over k_i^\perp} \Bigg)
  \nonumber\\
  &\times
  \bigintsss \prod\limits_{a=6}^{n-1} {d\sigma_a d\tau_a \over \tau_a}\,
  \bigintsss {d\rho d\tau_x \over \tau_x}\, \bigintsss {d\xi d\tau_y \over \xi}\,{\tau_x \over \tau_4 \tau_5}\,
  {\prod\limits_{I\in{\mathfrak N}} \delta^2\big( {S}^{\dot\alpha}_I \big) \prod\limits_{i\in\mathfrak{P}} \delta^2\big( {S}^\alpha_i \big) 
  \over (\rho {-} \sigma_6) \sigma_{67} \cdots\sigma_{n-2, n-1}\sigma_{n-1}},
\esp\eeq
where the scattering equations are given below in detail.
First, $\bar{S}^{\dot\alpha}_I$ become
\begin{align}\label{Collinear-SEs-I}
  \bar{S}^{\dot 1}_I \,&=\,
      {k_I^\perp}  - \sum_{i\in\mathfrak{P}\backslash\{4,5\}} {\tau_I \tau_i \over \sigma_I {-} \sigma_i} {k_i^+} 
      -  {\tau_I \tau_x \over \sigma_I {-} \rho} {k_x^+} 
      + {\tau_I \over 1 - \sum_{J\in{\mathfrak N}} t_J}\, q_4^\perp
       + {\cal O}(\epsilon),
  \\
  \bar{S}^{\dot 2}_I \,&=\,
      (k_I^\perp)^\ast  - {k_I^+ \over k_I^\perp}\sum_{i\in\mathfrak{P}\backslash\{4,5\}} {\tau_I \tau_i \over \sigma_I {-} \sigma_i} ({k_i^\perp})^\ast 
      - {k_I^+ \over k_I^\perp} {\tau_I \tau_x \over \sigma_I {-} \rho} ({k_x^\perp})^\ast
      - {\zeta_I  \over  1 + \sum_{J\in{\mathfrak N}} \zeta_J}\, ({q_n^\perp})^\ast
      + {\cal O}(\epsilon),
\nonumber
\end{align}
up to leading order in the collinear limit.
While ${S}^{\alpha}_i$ with $i\neq4,5$ remain unchanged, ${S}^{\alpha}_4$ and ${S}^{\alpha}_5$ become 
\begin{align}\label{}
  {S}^1_4 \,&=\, 1 +  \tau_4
  \left[1 - \sum_{I\in{\mathfrak N}}\left({\tau_I \over \rho {-} \sigma_I} - {\epsilon\xi \over 2} {\tau_I \over (\rho {-} \sigma_I)^2}\right) {k_I^+  \over  k_I^\perp}\right] 
  + {\cal O}(\epsilon^2),
  \\
  {S}^1_5 \,&=\, 1 + \tau_5
  \left[1 - \sum_{I\in{\mathfrak N}}\left({\tau_I \over \rho {-} \sigma_I} + {\epsilon\xi \over 2} {\tau_I \over (\rho {-} \sigma_I)^2}\right){k_I^+  \over  k_I^\perp}\right]
  + {\cal O}(\epsilon^2),
  \\
  {S}^2_4 \,&=\, 1 + 
  \tau_4 {k_x^+ \over k_x^\perp}   \Bigg\{
  {1 \over \rho}  - \sum_{I\in{\mathfrak N}}{\tau_I \over \rho {-} \sigma_I} 
  \nonumber\\
  &\qquad
  + \Bigg[
  \sqrt{1{-}z \over z}\, {\braket{xy} \over k_x^\perp} \bigg(  {1 \over \rho}  - \sum_{I\in{\mathfrak N}}{\tau_I \over \rho {-} \sigma_I} \bigg)
  - {\xi \over 2} \bigg({1 \over \rho^2} - \sum_{I\in{\mathfrak N}}  {\tau_I \over (\rho {-} \sigma_I)^2} \bigg)
  \Bigg]\epsilon \Bigg\} + {\cal O}(\epsilon^2),
  \\
  {S}^2_5 \,&=\, 1 + 
  \tau_5 {k_x^+ \over k_x^\perp} 
  \Bigg\{
  {1 \over \rho}  - \sum_{I\in{\mathfrak N}} {\tau_I \over \rho {-} \sigma_I}
  \nonumber\\
  &\qquad
  - \Bigg[\sqrt{z \over 1{-}z}\, {\braket{xy} \over k_x^\perp} \bigg( {1 \over \rho}  - \sum_{I\in{\mathfrak N}} {\tau_I \over \rho {-} \sigma_I} \bigg)
  - {\xi \over 2} \bigg( {1 \over \rho^2}  - \sum_{I\in{\mathfrak N}} {\tau_I \over (\rho {-} \sigma_I)^2} \bigg)
  \Bigg]\epsilon
  \Bigg\}
  + {\cal O}(\epsilon^2).
\end{align}
Next, we consider the following linear combinations,
\begin{align}
  \label{SE-collinear-pp-x}
  {S}^\alpha_x \,&\equiv\, z\,{S}^\alpha_4 + (1{-}z)\,{S}^\alpha_5,
  \\
  \label{SE-collinear-pp-y}
  {S}^\alpha_y \,&\equiv\, \sqrt{z(1{-}z)}\, \big( {-} {S}^\alpha_4 + {S}^\alpha_5\big).
\end{align}
such that $\delta^2\big({S}^\alpha_4\big) \delta^2\big({S}^\alpha_5\big)
  \,=\, z(1{-}z)\, \delta^2\big({S}^\alpha_x\big) \delta^2\big({S}^\alpha_y\big)$.
At leading order eq.~\eqref{SE-collinear-pp-x} reduces to
\begin{align}\label{Collinear-SEs-pp-x}
\begin{aligned} 
  {S}^1_x \,&=\, 1 +  \tau_x \left(1 - \sum_{I\in{\mathfrak N}} {\tau_I \over \rho {-} \sigma_I}\,{k_I^+  \over  k_I^\perp}\right) + {\cal O}(\epsilon),
  \\
  {S}^2_x \,&=\, 1  + \tau_x\left({1 \over \rho} - \sum_{I\in{\mathfrak N}} {\tau_I \over \rho {-} \sigma_I} \right) {k_x^+ \over k_x^\perp} + {\cal O}(\epsilon).
\end{aligned}
\end{align}
The interpretation of these equations is as follows: the collinear momenta have been replaced by a new parent particle with momentum $k_x$ and positive helicity, and to this particle we associate the variables ($\rho,\tau_x)$.
The remaining step is then to integrate out the variables $(\xi,\tau_y)$ associated to the collinear splitting using the equations ${S}^\alpha_y=0$ in eq.~\eqref{SE-collinear-pp-y}.

To leading order in $\eps$ the equations ${S}^\alpha_y=0$ are independent of $\xi$, so we need to expand them to next-to-leading order,
\begin{align}\label{}
  \label{collinear-pp-Sy-1}
  {S}^1_y \,=\, \tau_y\Lambda_1 - {1\over 2}&\epsilon\,\xi\Big( 2\sqrt{z(1{-}z)}\tau_x + (2z{-}1)\tau_y \Big)\Lambda_2 + {\cal O}(\epsilon^2),
  \\
  \label{collinear-pp-Sy-2}
  {S}^2_y \,=\,  {k_x^+ \over k_x^\perp}
  \Bigg\{
  \Lambda_3 \tau_y &- \bigg(\tau_x - {1-2z \over \sqrt{z(1{-}z)}} \tau_y\bigg) {\braket{x\,y} \over k_x^\perp} \Lambda_3 \epsilon
  \nonumber
  \\
  &+ {\xi \over 2} \Big( 2\sqrt{z(1{-}z)} \tau_x + (2z{-}1)\tau_y \Big) \Lambda_4 \epsilon
  \Bigg\} + {\cal O}(\epsilon^2),
\end{align}
where we introduced the shorthands,
\begin{align}
\begin{aligned}
  \Lambda_1 \,&=\, 1 - \sum_{I\in{\mathfrak N}} {\tau_I \over \rho {-} \sigma_I}\,{k_I^+  \over  k_I^\perp}, \\
  \Lambda_3 \,&=\,  {1 \over \rho} - \sum_{I\in{\mathfrak N}} {\tau_I \over \rho {-} \sigma_I}, \qquad
\end{aligned}
\qquad
\begin{aligned}
  \Lambda_2 \,&=\,  \sum_{I\in{\mathfrak N}} {\tau_I \over (\rho {-} \sigma_I)^2}\,{k_I^+  \over  k_I^\perp},   \\
  \Lambda_4 \,&=\,  {1 \over \rho^2} - \sum_{I\in{\mathfrak N}} {\tau_I \over (\rho {-} \sigma_I)^2}.
\end{aligned}
\end{align}
We use eq.\,\eqref{collinear-pp-Sy-1} to fix $\tau_y$, 
\begin{align}\label{}
  \tau_y \,&=\, {\epsilon \xi \sqrt{z(1{-}z)}\, \Lambda_2\,\tau_x  \over  \Lambda_1 - {1\over 2}\epsilon \xi (2z{-}1) \Lambda_2}
\end{align}
Finaly, we use the global residue theorem to express the residue at ${S}^2_y$ in terms of the residues at $\xi=0$ and $\xi=\infty$. The residue at $\xi=\infty$ vanishes, and we obtain
\beq\bsp
  V\big(q_4; 4^+,5^+,  \ldots, n{-}1; q_n\big)
&  \,\simeq\, {1 \over \epsilon\braket{x\,y}}\,{1 \over \sqrt{z (1 {-} z)}}\,
  V\big(q_4; K^+,6, \ldots, n{-}1; q_n\big)\\
&\,  =  \textrm{Split}_{-}(4^+,5^+)\,V\big(q_4; K^+,6, \ldots, n{-}1; q_n\big)\,.
\esp\eeq

\subsubsection*{$(h_4,h_5)=(+,-)$}
In the case where the collinear particles have opposite helicities, the general philosophy is similar to the previous case, though some of the steps are technically more involved. We only highlight the main steps here.
We perform the following change of variables,
\begin{align}
\begin{aligned}
  \tau_x \,&=\, z\,\tau_4 + \tau_5     \textrm{~~and~~}\tau_y \,&=\, \sqrt{z \over 1 {-} z}\,\big(\tau_5 - (1{-}z)\tau_4 \big)\,,
\end{aligned}
\end{align}
such that $d\tau_4 d\tau_5 \,=\, \sqrt{1 {-} z \over z}\, d\tau_x d\tau_y$, and eq.~\eqref{Lipatov-collinear-main} reduces to
\begin{align}\label{}
  V\big(q_4; 4,5, &\ldots, n{-}1; q_n\big)
  \,\simeq\, \big({q_4^\perp}^\ast {q_n^\perp}\big)
  \Bigg( \prod_{\substack{i\in\mathfrak{P}\backslash\{4\}\\ I\in{\mathfrak N}\backslash\{5\}}} {k_I^\perp \over k_i^\perp}  \Bigg)
  \left({1 {-} z \over z}\right)^{3/2}
  \nonumber\\
  &\times
  \bigintsss \prod\limits_{a=6}^{n-1} {d\sigma_a d\tau_a \over \tau_a}\,
  \bigintsss {d\rho d\tau_x} \bigintsss {d\xi d\tau_y \over \xi\, \tau_4 \tau_5}\,
  {\prod\limits_{I\in{\mathfrak N}} \delta^2\big( {S}^{\dot\alpha}_I \big) \prod\limits_{i\in\mathfrak{P}} \delta^2\big( {S}^\alpha_i \big) 
  \over (\rho {-} \sigma_6) \sigma_{67} \cdots\sigma_{n-2, n-1}\sigma_{n-1}}.
  \label{Lipatov-collinear-pm-1}
\end{align}
The scattering equations entering eq.~\eqref{Lipatov-collinear-pm-1} can be cast in the form:
\begin{align}\label{}
  {S}^{1}_i \,&=\,1 + \tau_i  - \sum_{I\in{\mathfrak N}\backslash\{5\}} {\tau_i \tau_I \over \sigma_i {-} \sigma_I} {k_I^+  \over  k_I^\perp}
  - {\tau_i \tau_5 \over \sigma_i {-} \rho} {k_x^+  \over  k_x^\perp}
  + {\cal O}(\epsilon),
  \\
  {S}^{2}_i \,&=\, 1  + {k_i^+ \over k_i^\perp} \bigg( {\tau_i \over \sigma_i} - \sum_{I\in{\mathfrak N}\backslash\{5\}} {\tau_i \tau_I \over \sigma_i {-} \sigma_I}
  - {\tau_i \tau_5 \over \sigma_i {-} \rho} \bigg)
  + {\cal O}(\epsilon),
  \\
  \bar{S}^{\dot 1}_I \,&=\,  {k_I^\perp}  - \sum_{i\in\mathfrak{P}\backslash\{4\}} {\tau_I \tau_i \over \sigma_I {-} \sigma_i} {k_i^+} 
  - {\tau_I \tau_4 \over \sigma_I {-} \rho}\, z{k_x^+} 
  + {a \tau_I}
  + {\cal O}(\epsilon),
  \\
  \label{collinera-SE-I2-1}
  \bar{S}^{\dot 2}_I \,&=\, (k_I^\perp)^\ast  -{k_I^+ \over k_I^\perp} \bigg(
  \sum_{i\in\mathfrak{P}\backslash\{4\}} {\tau_I \tau_i \over \sigma_I {-} \sigma_i} ({k_i^\perp})^\ast 
  + {\tau_I \tau_4 \over \sigma_I {-} \rho}\, z({k_x^\perp})^\ast 
  + b_0{\tau_I \over \sigma_I} \bigg)
  + {\cal O}(\epsilon),
  \\
  {S}^{1}_4 \,&=\,1 + \tau_4  - \sum_{I\in{\mathfrak N}\backslash\{5\}} {\tau_4 \tau_I \over \rho {-} \sigma_I {+} {1\over 2}\epsilon\xi} {k_I^+  \over  k_I^\perp}
  - {\tau_4 \tau_5 \over \epsilon\,\xi} {k_5^+  \over  k_5^\perp},
  \\
  {S}^{2}_4 \,&=\,1 + {k_4^+ \over k_4^\perp} \bigg(
  {\tau_4 \over \rho + \epsilon\,\xi/2} 
  - \sum_{I\in{\mathfrak N}\backslash\{5\}} {\tau_4 \tau_I \over \rho {-} \sigma_I {+} {1\over 2}\epsilon\xi}  -  {\tau_4 \tau_5 \over \epsilon\,\xi} \bigg),
  \\
  \bar{S}^{\dot 1}_5 \,&=\,  {k_5^\perp}  - \sum_{i\in\mathfrak{P}\backslash\{4\}} {\tau_5 \tau_i \over \rho {-} \sigma_i {-} {1\over 2}\epsilon\xi} {k_i^+} 
  + {\tau_5 \tau_4 \over \epsilon\,\xi} {k_4^+}
  + a \tau_5,
  \\
  \bar{S}^{\dot 2}_5 \,&=\, (k_5^\perp)^\ast  - {k_5^+ \over k_5^\perp} \bigg(
  \sum_{i\in\mathfrak{P}\backslash\{4\}} {\tau_5 \tau_i \over \rho {-} \sigma_i {-} {1\over 2}\epsilon\xi} ({k_i^\perp})^\ast 
  - {\tau_5 \tau_4 \over \epsilon\,\xi} ({k_4^\perp})^\ast
  + b {\tau_5 \over \rho {-} {1\over 2}\epsilon\xi}  \bigg),
\end{align}
with
\begin{align}\label{}
  a \,\equiv\, q_4^\perp \bigg(1 - \sum_{J\in{\mathfrak N}} t_J\bigg)^{-1} \textrm{~~and~~}
 b \,\equiv\, ({q_n^\perp})^\ast\bigg(1 + \sum_{J\in{\mathfrak N}} {t_J \over \sigma_J} {k_J^+  \over  k_J^\perp}\bigg)^{-1},
\end{align}
and $b_0$ in eq.~\eqref{collinera-SE-I2-1} is the leading order approximation of $b$ in $\epsilon$.

In the first step, we use $\delta({S}^1_4)$ to fix $\tau_y$.
Since ${S}^1_4=0$ is a quadratic equation in $\tau_y$, it has two solutions as follows:
\begin{align}
  \label{ty-sol-plus}
   \tau^{\pm}_y \,&=\,  {\big( (2z{-}1){k_5^+}\tau_x + \epsilon\xi\Xi{k_5^\perp}\big)
   \pm  \sqrt{-4z \epsilon\xi {k_5^\perp} {k_5^+} + \big({k_5^+}\tau_x  -  \epsilon\xi \Xi {k_5^\perp}\big)^2}
   \over  2{k_5^+} \sqrt{z(1{-}z)}}\,,
\end{align}
with
\begin{align}\label{}
  \Xi \,&\equiv\,1  - \sum_{I\in{\mathfrak N}\backslash\{5\}} {\tau_I \over \rho {-} \sigma_I {+} {1\over 2}\epsilon\xi} {k_I^+  \over  k_I^\perp}.
\end{align}
Equation~\eqref{Lipatov-collinear-pm-1} then reduces to
\begin{align}
  V\big(q_4; 4,5&, \ldots, n{-}1; q_n\big)
  \,\simeq\, ({q_4^\perp})^\ast {q_n^\perp}\,
  \Bigg( \prod_{\substack{i\in\mathfrak{P}\backslash\{4\}\\ I\in{\mathfrak N}\backslash\{5\}}} {k_I^\perp \over k_i^\perp}  \Bigg)
  (1-z)^{1/2} z^{-3/2}
  \nonumber\\
  &\times
  \bigintsss \prod\limits_{a=6}^{n-1} {d\sigma_a d\tau_a \over \tau_a}\,
  \bigintsss {d\rho d\tau_x \over \tau_x}
  \bigintsss {d\xi \over \xi}\, \Big( \Delta|_{\tau_y = t^{+}_y} - \Delta|_{\tau_y = t^{-}_y} \Big)
  {\tau_x \over {\tau_y^+} - {\tau_y^-}},
\end{align}
where
\begin{align}\label{collinear-Delta}
  \Delta  \,\equiv\,  {\delta\big({S}^2_4\big) \delta^2\big( \bar{S}^{\dot\alpha}_5 \big) \over (1+\tau_4\Xi)}\,
  {\prod\limits_{i\in\mathfrak{P}\backslash\{4\}}\delta^2\big( {S}^\alpha_i \big)
  \prod\limits_{I\in\mathfrak{N}\backslash\{5\}}\delta^2\big( \bar{S}^{\dot\alpha}_I \big)
  \over (\rho {-} \sigma_6) \sigma_{67} \cdots\sigma_{n-2, n-1}\sigma_{n-1}}.
\end{align}
Finally, we again use the global residue theorem and integrate out $\xi$ by summing the residues at $\xi=0$ and $\xi=\infty$. 
The residue at $\xi=\infty$ again vanishes, and we find
\begin{align}
  V\big(q_4; 4,5, \ldots, & n{-}1; q_n\big)
  \,=\, - ({q_4^\perp})^\ast {q_n^\perp}\,
  \Bigg( \prod_{i\in\mathfrak{P}\backslash\{4\}, I\in{\mathfrak N}\backslash\{5\}} {k_I^\perp \over k_i^\perp}  \Bigg)
  \left({1 {-} z \over z}\right)\,
  \nonumber\\
  &\times
  \bigintsss \prod\limits_{a=6}^{n-1} {d\sigma_a d\tau_a \over \tau_a}\,
  \bigintsss {d\rho d\tau_x \over \tau_x}\,
  \Big( \Delta|_{(\xi,\tau_y) = (0,\tau^{+}_y)} - \Delta|_{(\xi,\tau_y) = (0,\tau^{-}_y)} \Big)\,.
  \label{Lipatov-collinear-pm-2}
\end{align}

Let us analyse separately the contributions from the two terms in eq.~\eqref{Lipatov-collinear-pm-2}.
We start by considering the term containing $\Delta|_{(\xi,\tau_y) = (0,\tau^{+}_y)}$.
We have
\begin{align}
  \lim_{\xi \to 0} \tau^{+}_y \,&=\,  \tau_x\,\sqrt{z \over 1{-}z},
\end{align}
which implies $\tau_4=0$ and $\tau_5=\tau_x$. Inserting this into the scattering equations, we obtain
\begin{align}\label{}
  {S}^{1}_i \,&=\,1 + \tau_i  - \sum_{I\in{\mathfrak N}\backslash\{5\}} {\tau_i \tau_I \over \sigma_i {-} \sigma_I} {k_I^+  \over  k_I^\perp}
  - {\tau_i \tau_x \over \sigma_i {-} \rho} {k_x^+  \over  k_x^\perp}
  + {\cal O}(\epsilon),
  \\
  {S}^{2}_i \,&=\,1  + {k_i^+ \over k_i^\perp} \bigg(
  {\tau_i \over \sigma_i} - \sum_{I\in{\mathfrak N}\backslash\{5\}} {\tau_i \tau_I \over \sigma_i {-} \sigma_I}
  - {\tau_i \tau_x \over \sigma_i {-} \rho}
  \bigg)
  + {\cal O}(\epsilon),
  \\
  \bar{S}^{\dot 1}_I \,&=\,  {k_I^\perp}  - \sum_{i\in\mathfrak{P}\backslash\{4\}} {\tau_I \tau_i \over \sigma_I {-} \sigma_i} {k_i^+} 
  + {a' \tau_I}
  + {\cal O}(\epsilon),
  \\
  \bar{S}^{\dot 2}_I \,&=\, (k_I^\perp)^\ast  - {k_I^+ \over k_I^\perp}\bigg(
  \sum_{i\in\mathfrak{P}\backslash\{4\}} {\tau_I \tau_i \over \sigma_I {-} \sigma_i} ({k_i^\perp})^\ast 
  + b'\,{\tau_I \over \sigma_I}
  \bigg)
  + {\cal O}(\epsilon),
  \\
  \label{collinear-pm-S5-1}
  \bar{S}^{\dot 1}_5 \,&=\,  {k_x^\perp}  - \sum_{i\in\mathfrak{P}\backslash\{4\}} {\tau_x \tau_i \over \rho {-} \sigma_i} {k_i^+} 
  + a' \tau_x
  + {\cal O}(\epsilon)
  \\
  \label{collinear-pm-S5-2}
  \bar{S}^{\dot 2}_5 \,&=\, (k_x^\perp)^\ast  - {k_x^+ \over k_x^\perp}\bigg(
  \sum_{i\in\mathfrak{P}\backslash\{4\}} {\tau_x \tau_i \over \rho {-} \sigma_i} ({k_i^\perp})^\ast 
  + b'\, {\tau_x \over \rho}
  \bigg)
  + {\cal O}(\epsilon),
\end{align}
with $\zeta_x \,\equiv\, {\tau_x \over \rho} {k_x^+  \over  k_x^\perp}$ and
\begin{align}\label{}
  a' \,&=\,   q_2^\perp \bigg(1 - \tau_x - \sum_{J\in{\mathfrak N}\backslash\{5\}} t_J\bigg)^{-1}\,,\qquad
  b' \,=\,   {q_1^\perp}^\ast\bigg(1 + \zeta_x  + \sum_{J\in{\mathfrak N}\backslash\{5\}} \zeta_J \bigg)^{-1}\,.
\end{align}
In addition, in this case we have
\begin{align}
  1 + \tau_4\Xi \,=\, 1, \qquad
  {S}^{2}_4  \,&=\,  - {1 \over \sqrt{z(1{-}z)}}\,{\braket{x\,y} \over k_x^\perp}\epsilon + {\cal O}(\epsilon^2).
\end{align}
We see that $\bar{S}^{\dot\alpha}_5$ only depends on $k_x$, and not on $k_4$ and $k_5$ separately. Thus, if we let $\bar{S}^{\dot\alpha}_x = \bar{S}^{\dot\alpha}_5$, we obtain the scattering equations associated with a single parent particle with momentum $k_x$ and negative helicity, and we have
\beq\bsp
  -{1 \over \epsilon\braket{x\,y}}\,{(1 {-} z)^2 \over \sqrt{z(1 {-} z)}}\,
 & V\big(q_4; K^-,6, \ldots, n{-}1; q_n\big)\\
&    =  \textrm{Split}_{+}(4^+,5^-)\,  V\big(q_4; K^-,6, \ldots, n{-}1; q_n\big)\,.
\esp\eeq

We can perform a similar analysis for the term $\Delta|_{(\xi,\tau_y) = (0,\tau^{-}_y)}$, and letting ${S}^1_x = \bar{S}^{\dot 1}_5/{k_x^\perp}$ and $S^2_x = z\, {S}^{2}_4$ we obtain scattering equations for a parent parton with momentum $k_x$ and positive helicity. This gives a contribution
\beq\bsp
  {1 \over \epsilon\, [x\,y]}\,{z^2 \over \sqrt{z(1 {-} z)}}\,
 & V\big(q_4; K^+,6, \ldots, n{-}1; q_n\big)\\
    &\,=\,  \textrm{Split}_{-}(4^+,5^-)\,   V\big(q_4; K^+,6, \ldots, n{-}1; q_n\big)\,.
\esp\eeq

\subsection{Quasi Multi-Regge limits}
In this appendix we prove that our CHY-type formulas for the impact factors and the Lipatov vertices have the expected factorisation properties if the last particle, with momentum $k_{n{-}1}$, has much smaller rapidity as the other particles.
A similar analysis can be performed in the case where the first particle has much greater rapidity than the other.

Here we only discuss the impact factor $C(2; 3, \ldots, n{-}1)$ and derive the factorisation in the limit $y_3\simeq \cdots\simeq y_{n-2} \gg y_{n-1}$.
The argument follows the same lines as the derivation of the factorisation of the amplitude in QMRK in Section~\ref{sec:qmrk_C}, so we will be brief.
More precisely, we use the Conjecture 1 and two $\delta$-functions to localise the integrals over $\tau_{n-1}$ and $\zeta_{n-1}$.
The derivation depends on the helicity of the particle $n{-}1$, and we now discuss each case in turn.

Let us first study the case where the gluon with momentum $k_{n{-}1}$ has positive helicity.
The two equations ${S}_{n-1}^\alpha = 0$ are linear in $\zeta_{n-1}$ and $\tau_{n-1}$. More precisely, we have
\begin{align}\label{}
  {S}_{n-1}^1  \,&=\, 1 + \tau_{n-1}\left(1 + \sum_{I\in\overline{\mathfrak N}} \zeta_I \right) \,=\, 0,
  \qquad
  {S}_{n-1}^2  \,=\, 1 + \zeta_{n-1} \,=\, 0.
\end{align}
We can use these equations to fix $\zeta_{n-1}$ and $\tau_{n-1}$. We obtain,
\begin{align}\label{app-QMRK-sol}
  \zeta_{n-1}  \,&=\, -1 \textrm{~~and~~}
  \tau_{n-1} \,=\, - {1 \over 1 + \sum_{I\in\overline{\mathfrak N}} \zeta_I}.
\end{align}
We can insert eq.~\eqref{app-QMRK-sol} into the remaining scattering equations $\bar{S}_I^{\dot\alpha}$, and we get
\beq\bsp
  \bar{S}_I^1 \,&=\, {k_I^\perp}  - \sum_{i\in\mathfrak{P}\backslash\{n-1\}} {\tau_I \tau_i \over \sigma_I {-} \sigma_i} {k_i^+},
  \\
   \bar{S}_I^2  \,&=\, (k_I^\perp)^\ast  
   - {k_I^+ \over k_I^\perp} \sum_{i\in\mathfrak{P}\backslash\{n-1\}} {\tau_I \tau_i \over \sigma_I {-} \sigma_i} ({k_i^\perp})^\ast
  - \zeta_I\, {({q_n^\perp} - k_{n-1}^{\perp})^\ast \over  1 + \sum_{J\in\overline{\mathfrak N}} \zeta_J}.
\esp\eeq
After $\zeta_{n-1}$ and $\tau_{n-1}$ have been integrated out, we immediately find
\begin{align}\label{app-QMRK-fac}
  C\big(2^-, 3, \ldots, n{-}1\big) \,\simeq\, 
  C\big(2^-, 3, \ldots, n{-}2\big) \,{-1 \over |q_{n-1}^\perp|^2}\,
  V\big(q_{n-1}; n{-}1; q_n\big)\,,
\end{align}
where $q_{n-1} = q_n - k_{n-1}$, and the impact factor $C\big(2^-, 3, \ldots, n{-}2\big)$ in the right-hand side is given again by the CHY-type formula in eq.~\eqref{Impact-factor-main}.

Similarly, in another case where the gluon with momentum $k_{n{-}1}$ has negative helicity, we first use the equation
\begin{align}\label{}
   \bar{S}_{n-1}^{\dot 2}  \,&=\, (k_{n-1}^\perp)^\ast  - \zeta_{n-1}\, {(q_n^\perp)^\ast \over  1 + \sum_{J\in\overline{\mathfrak N}} \zeta_J} \,=\, 0
\end{align}
to localize the integral over $\zeta_{n-1}$, and the formula for the impact factor becomes
\begin{align}\label{}
  {q_n^\perp} \times & {-k_{n-1}^\perp\,(q_n^\perp)^\ast \over (k_{n-1}^\perp)^\ast (q_{n-1}^\perp)^\ast} \bigintsss
  \prod\limits_{a=3}^{n-2} {d\sigma_a d\tau_a \over \tau_a}\,
  {1 \over \sigma_{34}\cdots\sigma_{n-2}}\,
  \left( \prod_{i\in\mathfrak{P}, I\in\overline{\mathfrak N}\backslash\{n-1\}} {k_I^\perp \over k_i^\perp}  \right)
  \nonumber
  \\
  \times&\prod_{I\in\overline{\mathfrak N}\backslash\{n-1\}} 
  \delta\Bigg( {k_I^\perp}  - \sum_{i\in\mathfrak{P}} {\tau_I \tau_i \over \sigma_I {-} \sigma_i} {k_i^+} \Bigg)
  \delta\Bigg( (k_I^\perp)^\ast  - {k_I^+ \over k_I^\perp} \sum_{i\in\mathfrak{P}} {\tau_I \tau_i \over \sigma_I {-} \sigma_i} (k_i^\perp)^\ast
  - \zeta_I\, {({q_n^\perp} - k_{n-1}^{\perp})^\ast \over  1 + \!\!\!\sum\limits_{J\in\overline{\mathfrak N}\backslash\{n-1\}} \!\! \zeta_J} \Bigg)
  \nonumber
  \\
  \times&\bigintsss {d\tau_{n-1} \over \tau_{n-1}}\, \delta\Bigg( {k_{n-1}^\perp}  + \tau_{n-1} \sum_{i\in\mathfrak{P}} \zeta_i {k_i^\perp} \Bigg)
  \nonumber
  \\
  \times&\prod_{i\in\mathfrak{P}}
  \delta\Bigg( 1 + \tau_i  - \sum_{I\in\overline{\mathfrak N}} {\tau_i \tau_I \over \sigma_i {-} \sigma_I} {k_I^+  \over  k_I^\perp}   \Bigg)
  \delta\Bigg( 1 + \zeta_i  - {k_i^+ \over k_i^\perp}\sum_{I\in\overline{\mathfrak N}} {\tau_i \tau_I \over \sigma_i {-} \sigma_I}
  - \zeta_i\, \tau_{n-1} \Bigg).
\end{align}
Finally, we use the residue theorem to localise the integral over $\tau_{n-1}$ on the residues at $\tau_{n-1} = 0$ and  $\tau_{n-1} = \infty$. The residue at  $\tau_{n-1} = \infty$ vanishes, and the residue at  $\tau_{n-1} = 0$ immediately reproduces the desired factorisation formula.

\bibliography{SEMRK}

\providecommand{\href}[2]{#2}\begingroup\raggedright\begin{thebibliography}{10}

\bibitem{Cachazo:2013hca}
F.~Cachazo, S.~He and E.~Y. Yuan, \emph{{Scattering of Massless Particles in
  Arbitrary Dimensions}},
  \href{https://doi.org/10.1103/PhysRevLett.113.171601}{\emph{Phys. Rev. Lett.}
  {\bfseries 113} (2014) 171601}
  [\href{https://arxiv.org/abs/1307.2199}{{\ttfamily 1307.2199}}].

\bibitem{Cachazo:2013iea}
F.~Cachazo, S.~He and E.~Y. Yuan, \emph{{Scattering of Massless Particles:
  Scalars, Gluons and Gravitons}},
  \href{https://doi.org/10.1007/JHEP07(2014)033}{\emph{JHEP} {\bfseries 07}
  (2014) 033} [\href{https://arxiv.org/abs/1309.0885}{{\ttfamily 1309.0885}}].

\bibitem{Cachazo:2013iaa}
F.~Cachazo, S.~He and E.~Y. Yuan, \emph{{Scattering in Three Dimensions from
  Rational Maps}}, \href{https://doi.org/10.1007/JHEP10(2013)141}{\emph{JHEP}
  {\bfseries 10} (2013) 141} [\href{https://arxiv.org/abs/1306.2962}{{\ttfamily
  1306.2962}}].

\bibitem{Cachazo:2013gna}
F.~Cachazo, S.~He and E.~Y. Yuan, \emph{{Scattering equations and
  Kawai-Lewellen-Tye orthogonality}},
  \href{https://doi.org/10.1103/PhysRevD.90.065001}{\emph{Phys. Rev.}
  {\bfseries D90} (2014) 065001}
  [\href{https://arxiv.org/abs/1306.6575}{{\ttfamily 1306.6575}}].

\bibitem{Fairlie:1972zz}
D.~B. Fairlie and D.~E. Roberts, \emph{{Dual Models Without Tachyons\,--\,A New
  Approach}}, {\emph{unpublished Durham preprint [PRINT-72-2440]} (1972) }.

\bibitem{Roberts:1972ggn}
D.~E. Roberts, \emph{{Mathematical structure of dual amplitudes}}, Ph.D.
  thesis, Durham U., 1972.

\bibitem{Gross:1987kza}
D.~J. Gross and P.~F. Mende, \emph{{The High-Energy Behavior of String
  Scattering Amplitudes}},
  \href{https://doi.org/10.1016/0370-2693(87)90355-8}{\emph{Phys. Lett.}
  {\bfseries B197} (1987) 129}.

\bibitem{Gross:1987ar}
D.~J. Gross and P.~F. Mende, \emph{{String Theory Beyond the Planck Scale}},
  \href{https://doi.org/10.1016/0550-3213(88)90390-2}{\emph{Nucl. Phys.}
  {\bfseries B303} (1988) 407}.

\bibitem{Witten:2004cp}
E.~Witten, \emph{{Parity invariance for strings in twistor space}},
  \href{https://doi.org/10.4310/ATMP.2004.v8.n5.a1}{\emph{Adv. Theor. Math.
  Phys.} {\bfseries 8} (2004) 779}
  [\href{https://arxiv.org/abs/hep-th/0403199}{{\ttfamily hep-th/0403199}}].

\bibitem{Fairlie:2008dg}
D.~B. Fairlie, \emph{{A Coding of Real Null Four-Momenta into World-Sheet
  Coordinates}}, \href{https://doi.org/10.1155/2009/284689}{\emph{Adv. Math.
  Phys.} {\bfseries 2009} (2009) 284689}
  [\href{https://arxiv.org/abs/0805.2263}{{\ttfamily 0805.2263}}].

\bibitem{Dolan:2014ega}
L.~Dolan and P.~Goddard, \emph{{The Polynomial Form of the Scattering
  Equations}}, \href{https://doi.org/10.1007/JHEP07(2014)029}{\emph{JHEP}
  {\bfseries 07} (2014) 029} [\href{https://arxiv.org/abs/1402.7374}{{\ttfamily
  1402.7374}}].

\bibitem{Farrow:2018cqi}
J.~A. Farrow, \emph{{A Monte Carlo Approach to the 4D Scattering Equations}},
  \href{https://doi.org/10.1007/JHEP08(2018)085}{\emph{JHEP} {\bfseries 08}
  (2018) 085} [\href{https://arxiv.org/abs/1806.02732}{{\ttfamily
  1806.02732}}].

\bibitem{Liu:2018brz}
Z.~Liu and X.~Zhao, \emph{{Bootstrapping the Solutions of Scattering
  Equations}},  \href{https://arxiv.org/abs/1810.00384}{{\ttfamily
  1810.00384}}.

\bibitem{Schwab:2014xua}
B.~U.~W. Schwab and A.~Volovich, \emph{{Subleading Soft Theorem in Arbitrary
  Dimensions from Scattering Equations}},
  \href{https://doi.org/10.1103/PhysRevLett.113.101601}{\emph{Phys. Rev. Lett.}
  {\bfseries 113} (2014) 101601}
  [\href{https://arxiv.org/abs/1404.7749}{{\ttfamily 1404.7749}}].

\bibitem{Afkhami-Jeddi:2014fia}
N.~Afkhami-Jeddi, \emph{{Soft Graviton Theorem in Arbitrary Dimensions}},
  \href{https://arxiv.org/abs/1405.3533}{{\ttfamily 1405.3533}}.

\bibitem{Kalousios:2014uva}
C.~Kalousios and F.~Rojas, \emph{{Next to subleading soft-graviton theorem in
  arbitrary dimensions}},
  \href{https://doi.org/10.1007/JHEP01(2015)107}{\emph{JHEP} {\bfseries 01}
  (2015) 107} [\href{https://arxiv.org/abs/1407.5982}{{\ttfamily 1407.5982}}].

\bibitem{Zlotnikov:2014sva}
M.~Zlotnikov, \emph{{Sub-sub-leading soft-graviton theorem in arbitrary
  dimension}}, \href{https://doi.org/10.1007/JHEP10(2014)148}{\emph{JHEP}
  {\bfseries 10} (2014) 148} [\href{https://arxiv.org/abs/1407.5936}{{\ttfamily
  1407.5936}}].

\bibitem{Cachazo:2015ksa}
F.~Cachazo, S.~He and E.~Y. Yuan, \emph{{New Double Soft Emission Theorems}},
  \href{https://doi.org/10.1103/PhysRevD.92.065030}{\emph{Phys. Rev.}
  {\bfseries D92} (2015) 065030}
  [\href{https://arxiv.org/abs/1503.04816}{{\ttfamily 1503.04816}}].

\bibitem{Zlotnikov:2017ahq}
M.~Zlotnikov, \emph{{Leading multi-soft limits from scattering equations}},
  \href{https://doi.org/10.1007/JHEP10(2017)209}{\emph{JHEP} {\bfseries 10}
  (2017) 209} [\href{https://arxiv.org/abs/1708.05016}{{\ttfamily
  1708.05016}}].

\bibitem{Chakrabarti:2017zmh}
S.~Chakrabarti, S.~P. Kashyap, B.~Sahoo, A.~Sen and M.~Verma, \emph{{Testing
  Subleading Multiple Soft Graviton Theorem for CHY Prescription}},
  \href{https://doi.org/10.1007/JHEP01(2018)090}{\emph{JHEP} {\bfseries 01}
  (2018) 090} [\href{https://arxiv.org/abs/1709.07883}{{\ttfamily
  1709.07883}}].

\bibitem{Nandan:2016ohb}
D.~Nandan, J.~Plefka and W.~Wormsbecher, \emph{{Collinear limits beyond the
  leading order from the scattering equations}},
  \href{https://doi.org/10.1007/JHEP02(2017)038}{\emph{JHEP} {\bfseries 02}
  (2017) 038} [\href{https://arxiv.org/abs/1608.04730}{{\ttfamily
  1608.04730}}].

\bibitem{Kuraev:1976ge}
E.~A. Kuraev, L.~N. Lipatov and V.~S. Fadin, \emph{{Multi - Reggeon Processes
  in the Yang-Mills Theory}}, {\emph{Sov. Phys. JETP} {\bfseries 44} (1976)
  443}.

\bibitem{Lipatov:1976zz}
L.~N. Lipatov, \emph{{Reggeization of the Vector Meson and the Vacuum
  Singularity in Nonabelian Gauge Theories}}, {\emph{Sov. J. Nucl. Phys.}
  {\bfseries 23} (1976) 338}.

\bibitem{Lipatov:1991nf}
L.~N. Lipatov, \emph{{High-energy scattering in QCD and in quantum gravity and
  two-dimensional field theories}},
  \href{https://doi.org/10.1016/0550-3213(91)90512-V}{\emph{Nucl. Phys.}
  {\bfseries B365} (1991) 614}.

\bibitem{Fadin:1989kf}
V.~S. Fadin and L.~N. Lipatov, \emph{{High-Energy Production of Gluons in a
  QuasimultiRegge Kinematics}}, {\emph{JETP Lett.} {\bfseries 49} (1989) 352}.

\bibitem{DelDuca:1995zy}
V.~Del~Duca, \emph{{Equivalence of the Parke-Taylor and the
  Fadin-Kuraev-Lipatov amplitudes in the high-energy limit}},
  \href{https://doi.org/10.1103/PhysRevD.52.1527}{\emph{Phys. Rev.} {\bfseries
  D52} (1995) 1527} [\href{https://arxiv.org/abs/hep-ph/9503340}{{\ttfamily
  hep-ph/9503340}}].

\bibitem{He:2016vfi}
S.~He, Z.~Liu and J.-B. Wu, \emph{{Scattering Equations, Twistor-string
  Formulas and Double-soft Limits in Four Dimensions}},
  \href{https://doi.org/10.1007/JHEP07(2016)060}{\emph{JHEP} {\bfseries 07}
  (2016) 060} [\href{https://arxiv.org/abs/1604.02834}{{\ttfamily
  1604.02834}}].

\bibitem{Geyer:2014fka}
Y.~Geyer, A.~E. Lipstein and L.~J. Mason, \emph{{Ambitwistor Strings in Four
  Dimensions}},
  \href{https://doi.org/10.1103/PhysRevLett.113.081602}{\emph{Phys. Rev. Lett.}
  {\bfseries 113} (2014) 081602}
  [\href{https://arxiv.org/abs/1404.6219}{{\ttfamily 1404.6219}}].

\bibitem{DelDuca:1999iql}
V.~Del~Duca, A.~Frizzo and F.~Maltoni, \emph{{Factorization of tree QCD
  amplitudes in the high-energy limit and in the collinear limit}},
  \href{https://doi.org/10.1016/S0550-3213(99)00657-4}{\emph{Nucl. Phys.}
  {\bfseries B568} (2000) 211}
  [\href{https://arxiv.org/abs/hep-ph/9909464}{{\ttfamily hep-ph/9909464}}].

\bibitem{Antonov:2004hh}
E.~N. Antonov, L.~N. Lipatov, E.~A. Kuraev and I.~O. Cherednikov,
  \emph{{Feynman rules for effective Regge action}},
  \href{https://doi.org/10.1016/j.nuclphysb.2005.05.013,
  10.1016/j.nuclphysb.2005.013}{\emph{Nucl. Phys.} {\bfseries B721} (2005) 111}
  [\href{https://arxiv.org/abs/hep-ph/0411185}{{\ttfamily hep-ph/0411185}}].

\bibitem{Lipatov:1995pn}
L.~N. Lipatov, \emph{{Gauge invariant effective action for high-energy
  processes in QCD}},
  \href{https://doi.org/10.1016/0550-3213(95)00390-E}{\emph{Nucl. Phys.}
  {\bfseries B452} (1995) 369}
  [\href{https://arxiv.org/abs/hep-ph/9502308}{{\ttfamily hep-ph/9502308}}].

\bibitem{Duhr:2009uxa}
C.~Duhr, \emph{{New techniques in QCD}}, Ph.D. thesis, Universit\'{e}
  catholique de Louvain, 2009.

\bibitem{Prygarin:2011gd}
A.~Prygarin, M.~Spradlin, C.~Vergu and A.~Volovich, \emph{{All Two-Loop MHV
  Amplitudes in Multi-Regge Kinematics From Applied Symbology}},
  \href{https://doi.org/10.1103/PhysRevD.85.085019}{\emph{Phys. Rev.}
  {\bfseries D85} (2012) 085019}
  [\href{https://arxiv.org/abs/1112.6365}{{\ttfamily 1112.6365}}].

\bibitem{Weinzierl:2014vwa}
S.~Weinzierl, \emph{{On the solutions of the scattering equations}},
  \href{https://doi.org/10.1007/JHEP04(2014)092}{\emph{JHEP} {\bfseries 04}
  (2014) 092} [\href{https://arxiv.org/abs/1402.2516}{{\ttfamily 1402.2516}}].

\bibitem{Kalousios:2013eca}
C.~Kalousios, \emph{{Massless scattering at special kinematics as Jacobi
  polynomials}}, \href{https://doi.org/10.1088/1751-8113/47/21/215402}{\emph{J.
  Phys.} {\bfseries A47} (2014) 215402}
  [\href{https://arxiv.org/abs/1312.7743}{{\ttfamily 1312.7743}}].

\bibitem{DelDuca:1995ki}
V.~Del~Duca, \emph{{Real next-to-leading corrections to the multi\,--\,gluon
  amplitudes in the helicity formalism}},
  \href{https://doi.org/10.1103/PhysRevD.54.989}{\emph{Phys. Rev.} {\bfseries
  D54} (1996) 989} [\href{https://arxiv.org/abs/hep-ph/9601211}{{\ttfamily
  hep-ph/9601211}}].

\bibitem{Berends:1988zn}
F.~A. Berends and W.~T. Giele, \emph{{Multiple Soft Gluon Radiation in Parton
  Processes}}, \href{https://doi.org/10.1016/0550-3213(89)90398-2}{\emph{Nucl.
  Phys.} {\bfseries B313} (1989) 595}.

\bibitem{He:2016dol}
S.~He and Y.~Zhang, \emph{{Connected formulas for amplitudes in standard
  model}}, \href{https://doi.org/10.1007/JHEP03(2017)093}{\emph{JHEP}
  {\bfseries 03} (2017) 093}
  [\href{https://arxiv.org/abs/1607.02843}{{\ttfamily 1607.02843}}].

\bibitem{DelDuca:1996nom}
V.~Del~Duca, \emph{{Quark - anti-quark contribution to the multi - gluon
  amplitudes in the helicity formalism}},
  \href{https://doi.org/10.1103/PhysRevD.54.4474}{\emph{Phys. Rev.} {\bfseries
  D54} (1996) 4474} [\href{https://arxiv.org/abs/hep-ph/9604250}{{\ttfamily
  hep-ph/9604250}}].

\bibitem{Lipatov:1982vv}
L.~N. Lipatov, \emph{{Graviton Reggeization}},
  \href{https://doi.org/10.1016/0370-2693(82)90156-3}{\emph{Phys. Lett.}
  {\bfseries 116B} (1982) 411}.

\bibitem{zhengwen_gravity}
Z.~Liu{\emph{{ in preparation}} (2018) }.

\bibitem{Spradlin:2009qr}
M.~Spradlin and A.~Volovich, \emph{{From Twistor String Theory To Recursion
  Relations}}, \href{https://doi.org/10.1103/PhysRevD.80.085022}{\emph{Phys.
  Rev.} {\bfseries D80} (2009) 085022}
  [\href{https://arxiv.org/abs/0909.0229}{{\ttfamily 0909.0229}}].

\bibitem{Petersen:2015Eu}
T.~K. Petersen, \emph{{Eulerian Numbers}}. Springer, 2015.

\bibitem{Stieberger:2015kia}
S.~Stieberger and T.~R. Taylor, \emph{{Subleading terms in the collinear limit
  of Yang--Mills amplitudes}},
  \href{https://doi.org/10.1016/j.physletb.2015.09.075}{\emph{Phys. Lett.}
  {\bfseries B750} (2015) 587}
  [\href{https://arxiv.org/abs/1508.01116}{{\ttfamily 1508.01116}}].

\end{thebibliography}\endgroup

\end{document}